\documentclass{jkas}
\usepackage{amsmath}
\usepackage{bm,pseudocode,fancybox}
\usepackage{graphicx}
\usepackage{subfigure,color}
\usepackage{ulem} 

\usepackage[
colorlinks=true, 
linkcolor=blue, 
anchorcolor=blue,
citecolor=blue, 
filecolor=blue, 
menucolor=blue,
urlcolor=blue]{hyperref} 

\def\year{2015} 
\def\month{October} 
\def\volume{00} 
\def\issue{2} 
\def\beginpage{00} 
\def\endpage{00} 
\def\received{--, 2015} 
\def\accepted{--, 2015} 
\date{Received \received; accepted \accepted}

\newcommand{\algo}[1]{\texttt{#1}}

\providecommand{\abs}[1]{\lvert#1\rvert}

\def\aap{A\&A}

\def\apjs{ApJSS}

\title{ Horizon Run 4 Simulation: Coupled Evolution of Galaxies and Large-scale Structures of the Universe}

\author[a]{Juhan~Kim}
\author[b]{Changbom~Park}
\author[b,*]{Benjamin~L'Huillier}
\author[b]{Sungwook E.~Hong}

\affil[a]{  Center  for  Advanced  Computation,  Korea  Institute  for
  Advanced Study, 85 Hoegiro, Dongdaemun-gu, Seoul 130-722, Korea } 
\affil[b]{ School of  Physics, Korea Institute for  Advanced Study, 85
  Hoegiro, Dongdaemun-gu, Seoul 130-722, Korea } 

\def\email{
lhuillier@kias.re.kr
}

\def\corrauthor{
B. L'Huillier; e-mail:\texttt{\email}
}

\def\runningauthor{
J. Kim et al.
}

\def\runningtitle{
Horizon Run 4
}

\def\keywords{
cosmology: large-scale  structure of universe ---  galaxies: evolution
--- methods: numerical 
}
 
\def\abstracttext{
The Horizon Run  4 is a cosmological $N$-body  simulation designed for
the  study  of  coupled  evolution between  galaxies  and  large-scale
structures  of the  Universe, and  for  the test  of galaxy  formation
models. 
Using $6300^3$ gravitating particles in a  cubic box of $L_{\rm box} =
3150 ~h^{-1}{\rm Mpc}$,  we build a dense forest of  halo merger trees
to trace  the halo merger  history with  a halo mass  resolution scale
down to {$M_s = 2.7 \times 10^{11} h^{-1}{\rm M_\odot}$.} 
We build a set of particle and  halo data, which can serve as testbeds
for comparison of cosmological  models and gravitational theories with
observations. 
We find that the FoF halo  mass function shows a substantial deviation
from the universal form with  tangible redshift evolution of amplitude
and shape.  
At  higher   redshifts,  the  amplitude   of  the  mass   function  is
lower{,}  and the  functional  form is  shifted toward  larger
values of {$\ln (1/\sigma)$}. 
We also  find that  the baryonic acoustic  oscillation feature  in the
two-point correlation function of mock galaxies becomes broader with a
peak  position  moving  to  smaller  scales  and  the  peak  amplitude
decreasing  for increasing  directional cosine  $\mu$ compared  to the
linear predictions. 
From  the halo  merger  trees  built from  halo  data at  {75}
redshifts, we measure the half-mass epoch of halos 
and find that  less massive halos tend to reach  half of their current
mass at higher redshifts. 
{Simulation  outputs including  snapshot data,  past lightcone
  space   data,    and   halo    merger   data   are    available   at
  \url{http://sdss.kias.re.kr/astro/Horizon-Run4/}.} 
}

\begin{document}
\jkashead 

\section{Introduction}
Over  the  last decade,  {a  series  of cosmological  $N$-body
  simulations called} the Horizon Run (HRs) simulations have served as
a  testbed  for  cosmological  models  through  comparisons  with  the
observed large-scale distribution of galaxies. 
The first  Horizon Run (HR1)  was performed  in 2008 and  published in
2009 \citep{kim09}. 
The  simulation  box   size  was  $L_{\rm  box}   =  6592  ~h^{-1}{\rm
  Mpc}${,} and the number of  evolved particles was $N_{\rm p}
= 4120^3$.  
The initial power  spectrum was calculated by a  fitting function from
\citet{einsenstein98}, adopting a standard  $\Lambda$ cold dark matter
($\Lambda$CDM)   cosmology  in   a   concordance   with  WMAP   5-year
observations \citep{dunkley09}. 
It produced eight non-overlapping all-sky  lightcone data of halos and
subhalos up to $z=0.6$. 
We  studied  the  non-linear  gravitational effects  on  the  baryonic
acoustic oscillation (BAO) peak  by measuring the changes {in}
the peak position and amplitude.  
In 2011, we performed even bigger simulations called Horizon Run 2 and
3 (HR2 and HR3, respectively; \citealt{kim11}). 
By adopting the  same cosmological model as in HR1,  the initial power
spectra  of  HR2  and  HR3   were  generated  from  the  CAMB  package
\citep{lewis00}.  
The simulated galaxy distributions  have been extensively exploited to
measure both the  expected distribution of the  largest structures for
testing cosmic homogeneity \citep{park12}  and the cosmic topology for
constraining the non{-}linear gravitational effect on the halo density
field \citep{choi13,kim14,speare15}.  

All the  previous HRs  have mean particle  separations larger  than $1
~h^{-1}{\rm Mpc}${, which has been sufficient for many} {cosmological}
tests.  
With much success in  quantifying the non-linear gravitational effects
on large-scale structures, recently we  extended our research focus to
galaxy formation studies.  
To  model  galaxies in  simulations,  we  employed the  subhalo-galaxy
one-to-one correspondence model and abundance matching between subhalo
mass  function and  the  observed galaxy  luminosity  or stellar  mass
function \citep{kim08}.  
Most {characteristics}  of observed galaxy distributions  (in terms of
luminosity  functions and  one-point density  distributions) are  well
reproduced by the model while observed abundances of {galaxy clusters}
are not properly recovered from subhalos.  
The underpopulation of  simulated {galaxy clusters} may  come from the
inefficient subhalo findings  in cluster regions (\citealt{muldrew11};
  for the various subhalo  finding comparisons see \citealt{onions12})
  or from the spatial decoupling between subhalos and galaxies{.  The}
  latter may survive  the tidal disruption longer due  to more compact
  sizes through baryonic dissipation \citep{weinberg08}.

In the  $\Lambda$CDM cosmology, dark matter  halos form hierarchically
through the merger of smaller structures.  
These  merger  events can  trigger  star  formation and  drive  galaxy
formation and evolution \citep{kauffmann04, blanton07}. 
 {The  merger history  of galaxies}  has extensively  been studied  in
 semi-analytic  models (SAMs;  \citealt{cole94,kauffmann97, delucia04,
   baugh06, lee14}) {for} the last two decades.  
In {SAMs}, the  gas heating and cooling  rate are tabulated{,}
and the resulting  star formation and supernovae  feedback effects are
implemented with some parametric prescriptions. 
Those parameters are fine-tuned to reproduce the correlation functions
and/or luminosity  functions of  observed galaxies. Even  though {SAMs
  have  achieved} a  great  success in  reproducing some  observables,
{they require  the introduction of  a large number of  parameters that
  are not necessarily physical}. 

Another  well-known  empirical  galaxy  model,  the  {halo  occupation
  distribution (HOD;}  \citealt{berlind02,zheng05, zheng09}) modeling,
has been  adopted to match the  inner part of the  galaxy correlation,
which is attributed to satellite pairs inside a virialized halo.  
To distribute satellite galaxies in a halo{,} they empirically measure
he probability  number distribution  of satellites  from observations.
The  HOD is  simpler than  SAMs, and  widely used  for the  comparison
between observed galaxies and simulated halos.  

The   galaxy-subhalo  correspondence   (or  the   abundance  matching;
\citealt{kim08,trujillo-gomez11,rodriguez-puebla13,         reddick13,
  klypin15}) model is positioned between the two aforementioned models.  
It is much simpler than SAMs but based on more physical processes than
the HOD.  
It originally  models the  satellite galaxy distribution  from subhalo
catalogs.  
{Satellite galaxies} {in  a galaxy cluster} originally  formed {\it in
  situ} isolated {and} merged into {the} cluster {afterward}.  
{While}  falling  into  the  potential well  of  {the}  cluster,  they
experience    {a}   drag    force   by    dynamical   friction
\citep{zhao04}{,}  and they  inevitably show  spiraling inward
orbital motions.  
 {After a certain time,} they finally merge into the central galaxy. 

{Although it seems} reasonable to assume {the presence of} a satellite
galaxy  inside  a   subhalo  as  long  as  there   is  no  galaxy-halo
decoupling{,} it has  been noted that some satellite  galaxies may not
have a host subhalo \citep{gao04, guo11, guo14, wang14}. 
This  could  be  tested   by  extensive  hydrodynamical  simulations  to
investigate the  segregation between satellite {galaxies}  and {their}
dark matter host \citep{weinberg08}.  
However,  hydrodynamical simulations  are {expensive}  to run  and still
require much  effort to reduce ambiguities  in astrophysical processes
and numerical artifacts. 
On  the other  hand, \cite{hong15}  {recently} showed  that if  most
bound particles are {used} instead  of subhalos {in the modeling}, {it
  is possible to identify such satellites without hosting subhalo.}  

Therefore,           we            performed           a           new
{simulation} in our  series, the Horizon
Run 4 (HR4).  
This  simulation,  with improved  spatial  and  mass resolutions  with
respect to the previous runs, retains a large number of particles.  
It  is  well-suited to  study  galaxy  formation by  producing  merger
trees. 

The outline of the paper is as follows{.}
In Section 2 {and 3,} we describe the simulation specifics and outputs
{of HR4, respectively}. 
 {Mass function, shape and spin of virialized halos} are dealt
 with in Section 4. 
The analysis  of two-point  correlation functions and  {mass accretion
  history} are given in Section 5 and 6, respectively. 
Summary and discussions are following in Section 7.
 
\section{GOTPM Code and Simulation}

\subsection{Initial Conditions \& Parallelism}

The simulation  was run  with an  improved version  of the  GOTPM code
\citep{dubinski04}. 
The input  power spectrum is calculated  by the CAMB package,  and the
initial positions  and velocities of  the particles are  calculated by
applying the second-order Lagrangian perturbation theory (2LPT) method
proposed by \cite{jenkins10}.  
The gravitational  force is evaluated through  splitting the Newtonian
force law {into  long- and short-range forces} (for  the Newtonian and
Relativistic 
relations, see \citealt{rigopoulos15, hwang12}). 
The long-range  forces are calculated  {from} the Poisson  equation in
Fourier space {for}  the density mesh built by  the Particle-Mesh (PM)
method{.} 
{The} short-range {forces are} measured with the Tree method. 

We  parallelized the  GOTPM code  implementing MPI  and OpenMP  with a
one-dimensional domain decomposition ($z$-directional slabs). 
We  adopt a  dynamic domain  decomposition, which  sets the  number of
particle in each domain to be equal within one percent.  
Accordingly, the slab width changes during the simulation run. 
By using a  dynamic domain decomposition, one can  easily identify the
neighborhoods of a domain and establish communications between them.  
On the other hand, slab domains usually have greater surface-to-volume
ratios  than ordinary  cubic domains  (e.g., the  orthogonal recursive
bisection, \citealt{dubinski96}),  and so  it has  large communication
size between domains.

\subsection{Non-recursive Oct-Sibling Tree}
We have  employed a non{-}recursive  oct-sibling tree {(OST)}  for the
tree-force update. 
The {OST}  is a  structured tree  of particles  and nodes  with mutual
connections established by sibling and daughter pointers.  
Each particle has one sibling pointer, and each node has two pointers:
one for its daughter and the other for the next sibling.  

\begin{figure}[tp]
\centering
\includegraphics[width=8.4cm]{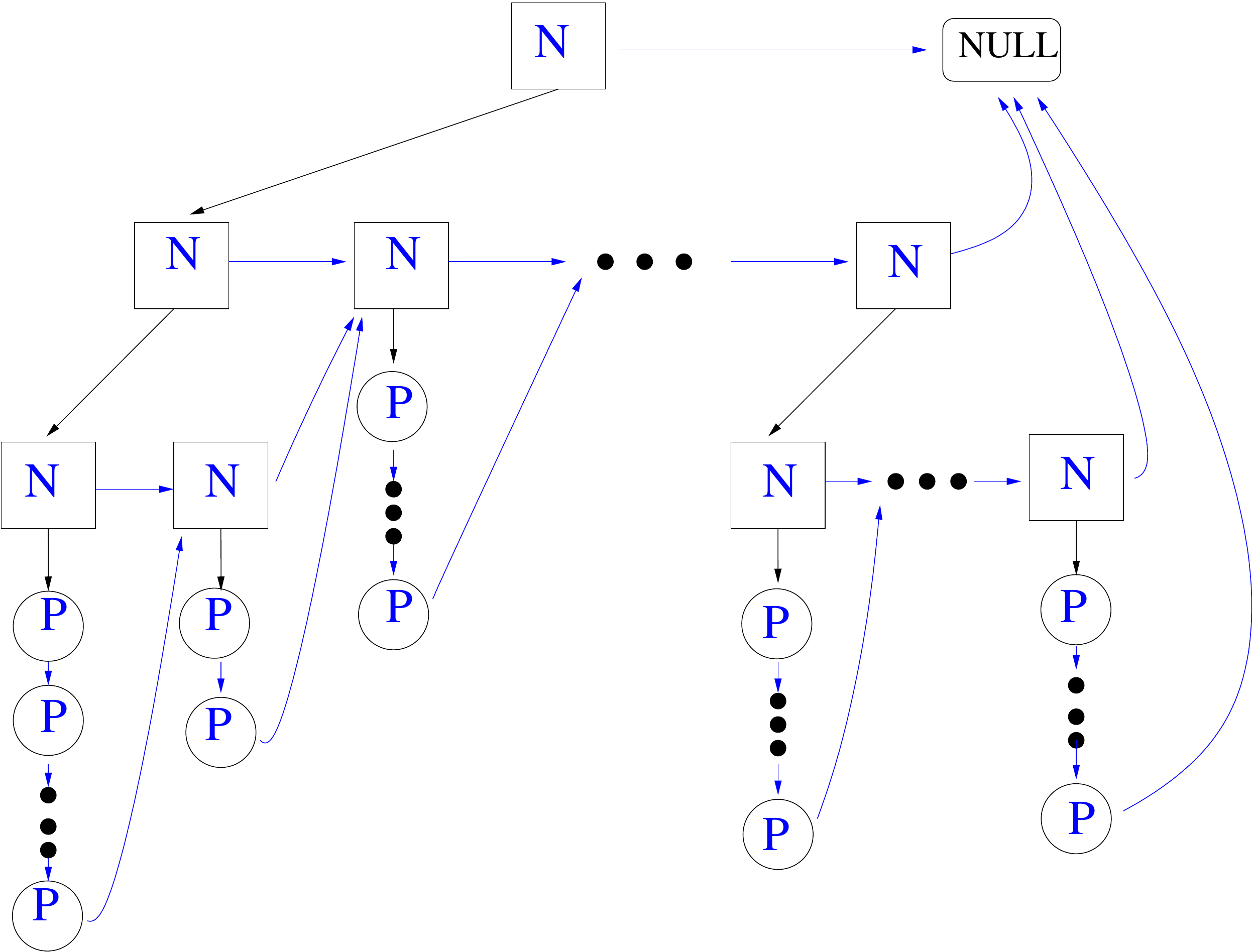}
\caption{Example of the Oct-Sibling  Tree structure. Boxes and circles
  represent  nodes and  particles,  respectively. The  black and  blue
  arrows are daughter and sibling pointers, respectively. 
Each node has a daughter and  sibling pointers while each particle has
only a sibling pointer.  
}
\label{ostfig}
\end{figure}
First, we create a top-most node encompassing all particles. 
We  define it  as the  zero  tree level,  and its  sibling pointer  is
directed to the null value (Fig. \ref{ostfig}). 
From the top-most  node, we recursively divide {each}  node into eight
equal-sized cubic subnodes by increasing the tree level by one. 
If a  sibling subnode contains  more {particle{s} than  a {pre}defined
  number}, we divide the node further  by increasing the tree level by
one. 
The daughter  pointer of  the node  is directed  to the  first sibling
subnode{,}  and  the  other   subnodes  are  linked  by  their
 sibling pointers.  
If the node  does not have any  subnode, we make a  chain of particles
linked by their sibling pointers, and we  set the start and end of the
chain connected  to the previous  and next sibling nodes  (or possibly
particles), respectively.  
The last sibling at  each local tree level is set  to have its sibling
pointer directed to the mother's next  sibling if it exits. If not, we
also recursively climb  the local tree until we find  the next sibling
of the current tree line. 


The advantage of the OST  over the traditional oct{-}tree is a
smaller  number of  pointers it  employs and  {the needlessness  of} a
recursive  tree  walk, which  requires  additional  costs for  such  a
stacking  process  to temporarily  store  information  of the  current
recursive depth. 
The algorithm \ref{ost} is a pseudocode for the non-recursive tree walk
with the OST. 
The tree walk is taken until  the running pointer, $p$, encounters the
null value{.}  
During  the tree  walks the  opening of  a node  is determined  by the
\algo{Open} function.  
The  tree-force   update  is  done  either   by  \algo{GroupForce}  or
\algo{ParticleForce}  depending  on the  data  type  addressed by  the
pointer ($p\rightarrow {\bf type}$). 
These three functions  play a pivotal role in  tree walks. \algo{Open}
decides whether to go further into  one deeper level (opening the node
and going down to its daughter) or  jump to the next sibling under the
opening  condition  that  $\theta  > \theta_c$,  with  $\theta_c$  the
predefined opening threshold.  
The \algo{GroupForce} function calculates the gravitational force from
the   group  of   particles  using   the  multipole   expansion  while
\algo{ParticleForce}  calculates  the  gravitational  force  from  the
particle, $p$. 
Thanks to  its cost efficiency, this kind of pseudocode  is {widely
  applied}  to  our  analysis  tools  such  as  a  percolation  method
(Friend-of-Friend  halo   finding),  peak  findings  in   a  Spherical
Overdensity  halo   identification,  and  the   two-point  correlation
measurement.

\begin{figure}
\begin{minipage}{1\columnwidth}
\begin{pseudocode}{GotpmTreeWalk}{p}
\WHILE p \neq \bf{NULL} \DO
\BEGIN
    \IF (p\rightarrow {\bf type}) = {\rm NODE} \THEN
    \BEGIN
        \IF {\bf Open}(p) = {\rm YES} \THEN
        \BEGIN
            p := p\rightarrow {\bf daughter}
        \END
        \ELSE 
        \BEGIN
            \boldsymbol{call}~~ {\bf GroupForce}(p)\\
            p := p\rightarrow {\bf sibling}
        \END
     \END
     \ELSE 
     \BEGIN
         \boldsymbol{call} ~~{\bf ParticleForce}(p)\\
          p := p\rightarrow {\bf sibling}
     \END
\END\\
\label{ost}
\COMMENT{$p$ is a running pointer.}
\end{pseudocode}
\end{minipage}
\end{figure}

\subsection{Position Accuracy in GOTPM}
One of the key factors to determine the resolution of Lagrangian codes
is  the  spatial  accuracy  or,   more  specifically,  the  number  of
significant digits involved in the particle position.  
Usually, a  single-precision floating-point  type has been  applied to
save the position of a particle because the four-byte single precision
is sufficient for small simulations.  
However,  as the  number  of particles  in  simulations increases,  the
position accuracy from single-precision begins to deteriorate.  
Since the roundoff error of a single precision variable $\mathcal{A}$
is    $\varepsilon_{\rm    roundoff}(\mathcal{A})    \sim    10^{-7}
\mathcal{A}$,  the maximum  roundoff error  of the  single-precision
position with respect to the mean particle separation is 
\begin{equation}
  \varepsilon_{\rm roundoff}\left(\frac{r_{\rm max}}{d_{\rm mean}} \right)
  \sim 10^{-7} \frac{L_{\rm box}}{L_{\rm box} / N_{\rm p}^{1/3}} 
  \sim 10^{-7} N_{\rm p}^{1/3}.
\end{equation}
For example, if the total number  of particles is $6300^3$ as in the
HR4, the maximum roundoff position error  lies at the level of a few
sub-percent  of the  mean  particle  separation, or  $\varepsilon_{\rm
  roundoff}(r_{\rm max} / d_{\rm mean}) \sim 10^{-3}$.  

On the other hand, in the HR4 as well as in the HR2 \& 3, we
  separate the  position of a  particle {($\boldsymbol r$)}  into two
vectors as 
 \begin{equation}
{\boldsymbol r} = {\boldsymbol L} + {\boldsymbol d},
\label{precision}
\end{equation}
where  $\boldsymbol  L$  {and  $\boldsymbol  d$  are}  the  Lagrangian
position {and  displacement from the {Lagrangian} position}
of  a particle{,  respectively}{.  We}  set the  particle  index by  a
row-major order  in the {Lagrangian} configuration  and, therefore, it
does not require additional memory space to compute $\boldsymbol L$.  
{Since the displacement of the  simulated particle over the entire HR4
  simulation  run  is  less  than  ten  times  of  the  mean  particle
  separation  ($d_{\rm max}  \lesssim 10  d_{\rm mean}$),  the maximum
  position error in the HR4 is 
\begin{equation}
\varepsilon_{\rm roundoff}\left( \frac{r_{\rm max}}{d_{\rm mean}} \right) \sim
10^{-7} \frac{d_{\rm max}}{d_{\rm mean}} \sim 10^{-6} .
\end{equation}
In this  way, we significantly  enhanced the accuracy of  the particle
position without using any additional memory space.  
}

\subsection{Simulation Specifics}
The HR4 was performed on  the supercomputer of Tachyon-II installed at
KISTI (Korea Institute of Science and Technology Information). 
We used  8,000 CPU cores over  50 straight days from  late November in
2013 to early February in 2014. 
Even with several  system glitches over the allocated  time period, we
succeeded  to   complete  the   simulation  in   about  50   days  for
the gravitational evolution of $6300^3$ particles in a periodic cubic box
of a side length $L_{\rm box} = 3150 ~h^{-1}{\rm Mpc}$. 
The starting  redshift is $z_i=100$,  which is chosen  for {particles}
not to overshoot one grid  cell spacing \citep{lukic07} in setting the
initial conditions.  
{This high initial  redshift, combined with 2LPT,  ensures an accurate
  power   spectrum   and   mass    function   measurement   at   $z=0$
  \citep{benjamin14}.}  
The simulation took 2000 steps to reach the final epoch of $z_0=0$. 
The mean particle separation is set to $d_{\rm mean} = 0.5 ~h^{-1}{\rm
  Mpc}$ and the corresponding force resolution is $0.1 d_{\rm mean}$. 

We adopted  a standard $\Lambda$CDM cosmology  in concordance
  with WMAP 5-year.
This  choice  of  cosmology  was made  for  consistency  with  various
observations including SDSS as well as the previous HRs. 
Specifically, the  matter, baryonic matter, and  dark energy densities
are     $\Omega_{m,0}    =     0.26$,    $\Omega_{b,0}=0.044$,     and
$\Omega_{\Lambda,0} = 0.74$, respectively.  
The current Hubble expansion is $H_0=100~h$ km/s/Mpc, where $h=0.72$. 
The amplitude  of the  initial matter perturbations  is scaled  for an
input bias factor, $b_8\equiv 1/\sigma_8=1.26$, where 
\begin{equation}
\sigma_8^2 = {1\over 2\pi^2} \int k^2 P(k) |W(kR_8)|^2 {\rm d}k,
\end{equation}
{and $R_8 \equiv 8 h^{-1}{\rm Mpc}$.}
Here,  {we used}  the spherical  top-hat filter  $W(x) \equiv  3(x\sin
x-\cos x)/x^3$ in $k$-space. 
{The  particle   mass  is  $m_{\rm  p}   \simeq  9\times  10^9
  ~h^{-1}{\rm M_\odot}$,  and }  the minimum mass  of halos  {with} 30
member particles  is about  $M_s \simeq 2.7\times  10^{11} ~h^{-1}{\rm
  M_\odot}$. 

\begin{figure}[tp]
\centering
\includegraphics[width=8.4cm]{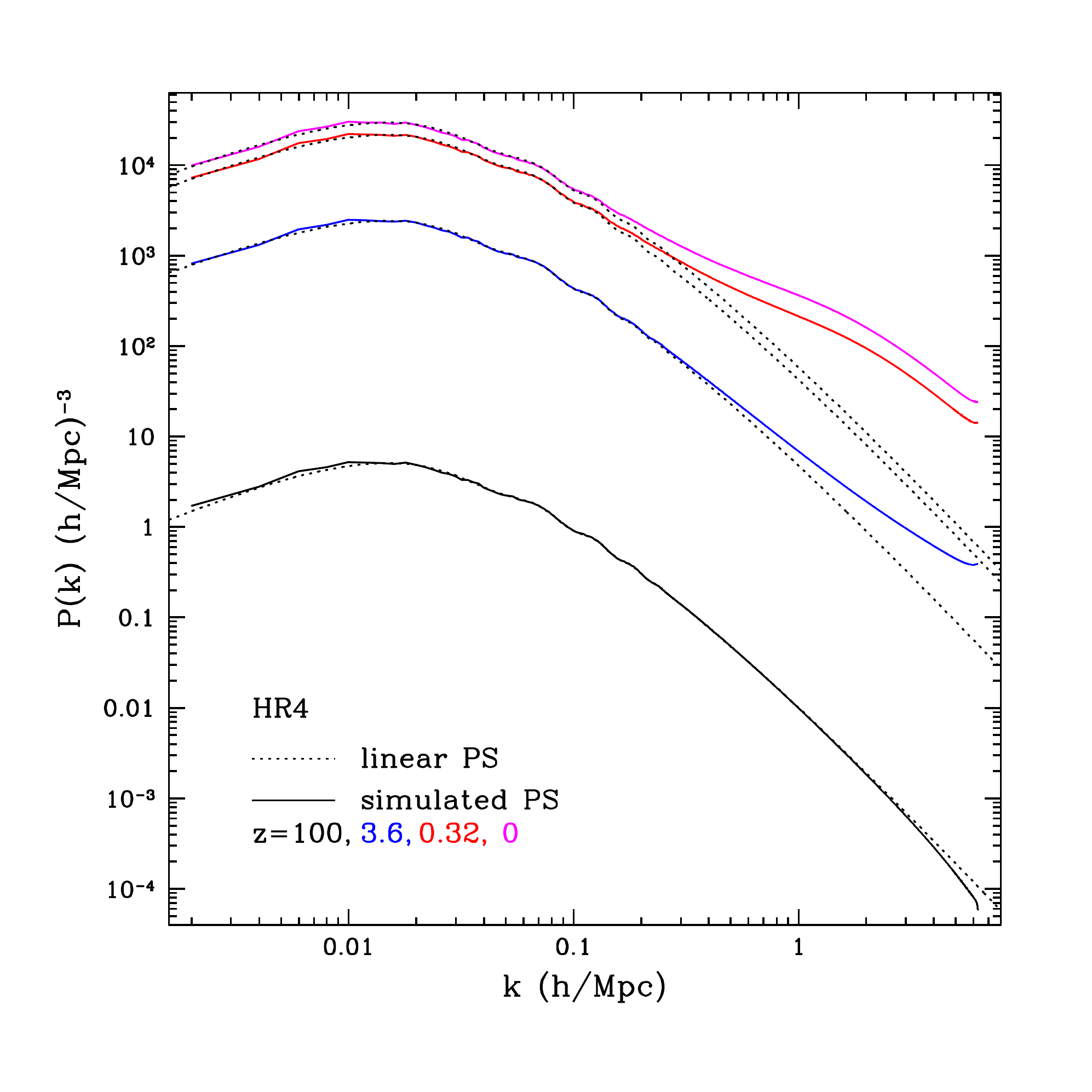}
\caption{Matter power 
{spectra from the HR4 simulation ({\it solid}) and from linear
  theory ({\it dotted}). 
{\it Color}:  $z = 100$  ({\it black}),  3.6 ({\it blue}),  0.32 ({\it
  red}), and 0 ({\it magenta}).} 
 \label{fig1} }
\end{figure}

Figure~\ref{fig1}  shows  the  evolution of  the  non{-}linear
matter power  spectrum obtained during  the simulation run  at several
redshifts. 
The dotted lines are the expected linear power spectra, while the solid
lines are the simulated matter power spectra at the same redshift. 
The typical  non-linear evolution effect  can easily be seen  on small
scales, where the  amplitude of the power spectrum  is greater than
the linear prediction due to the gravitational clustering.

\begin{figure*}[tp]
\centering
\includegraphics[width=17cm]{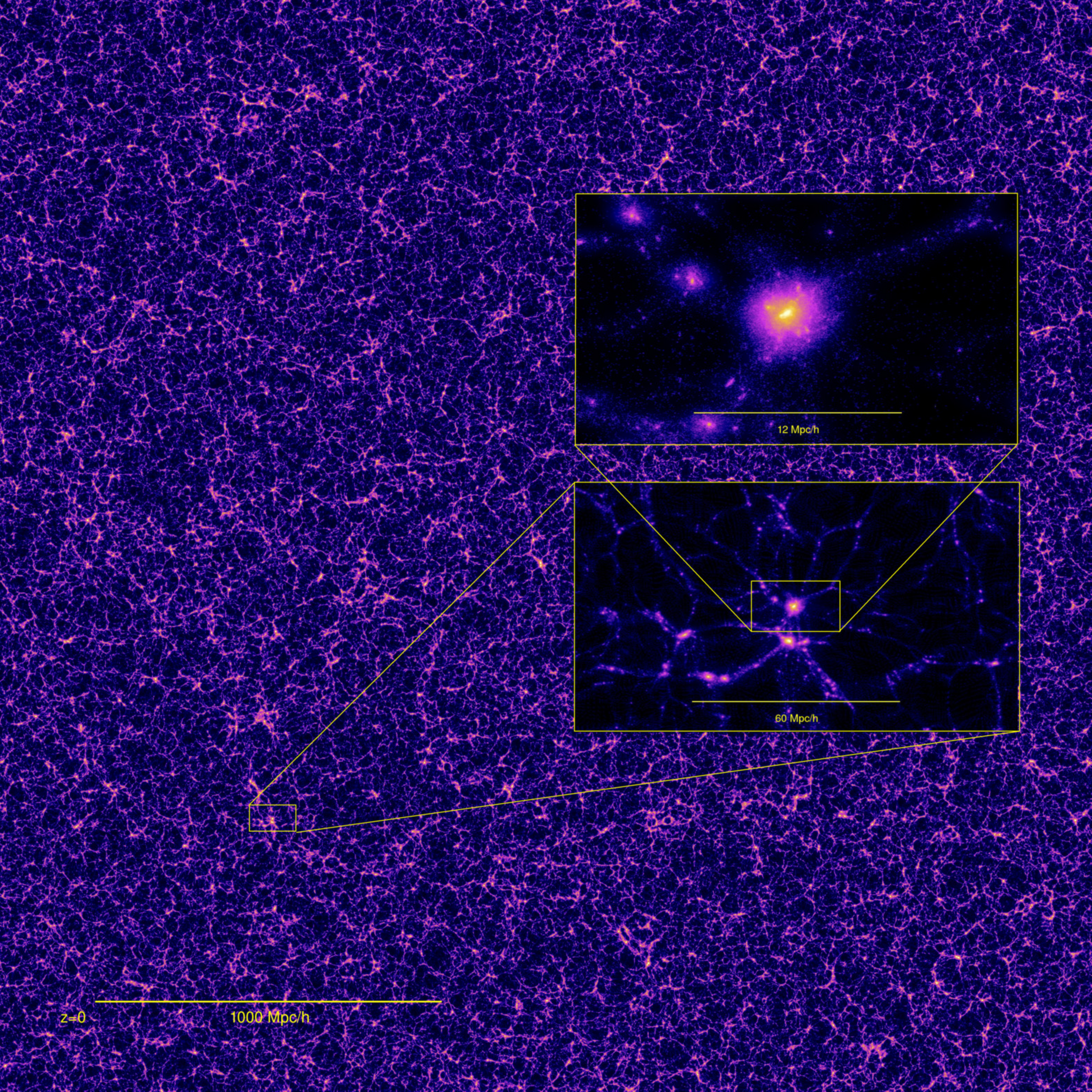}
\caption{Simulation density slice map at $z=0$.  
  High-density regions are {painted with bright color}.  
The width of the slice is $7 ~h^{-1}{\rm Mpc}$. 
The  {two} subfigures  are arranged  for cascaded  zoom-in views  of a
cluster at the center of the box in the bottom part of the figure.  
We put a scaling bar on the bottom of each panel.
 \label{HR4} }
\end{figure*}

Figure~\ref{HR4} shows a part of the density map of the HR4 at $z=0$,
where a cluster develops at the  center through the mergers of several
neighboring  overdensity  clumps. One  may  clearly  see void  regions
(painted in dark blue) with a size of a few tens of $h^{-1}$Mpc. 
Some  overdense  blobs are  embedded  in  the connection  of  multiple
filamentary structures.  

\section{Outputs}
{In this  section, we  describe the main  products of  the HR4
  simulation.          They          are         available          at
  \url{http://sdss.kias.re.kr/astro/Horizon-Run4/}.} 

\subsection{Snapshot and Past Lightcone Space Data}
We have saved snapshot data of particles at twelve redshifts: 
$z=0$, 0.05, 0.1, 0.15, 0.2, 0.3, 0.4, 0.5, 0.6, 0.7, 1, and 4. 
Each data set contains the particle position, velocity, and eight-byte
integer ID index.  


\begin{figure*}[t!]
\centering
\includegraphics[width=18cm]{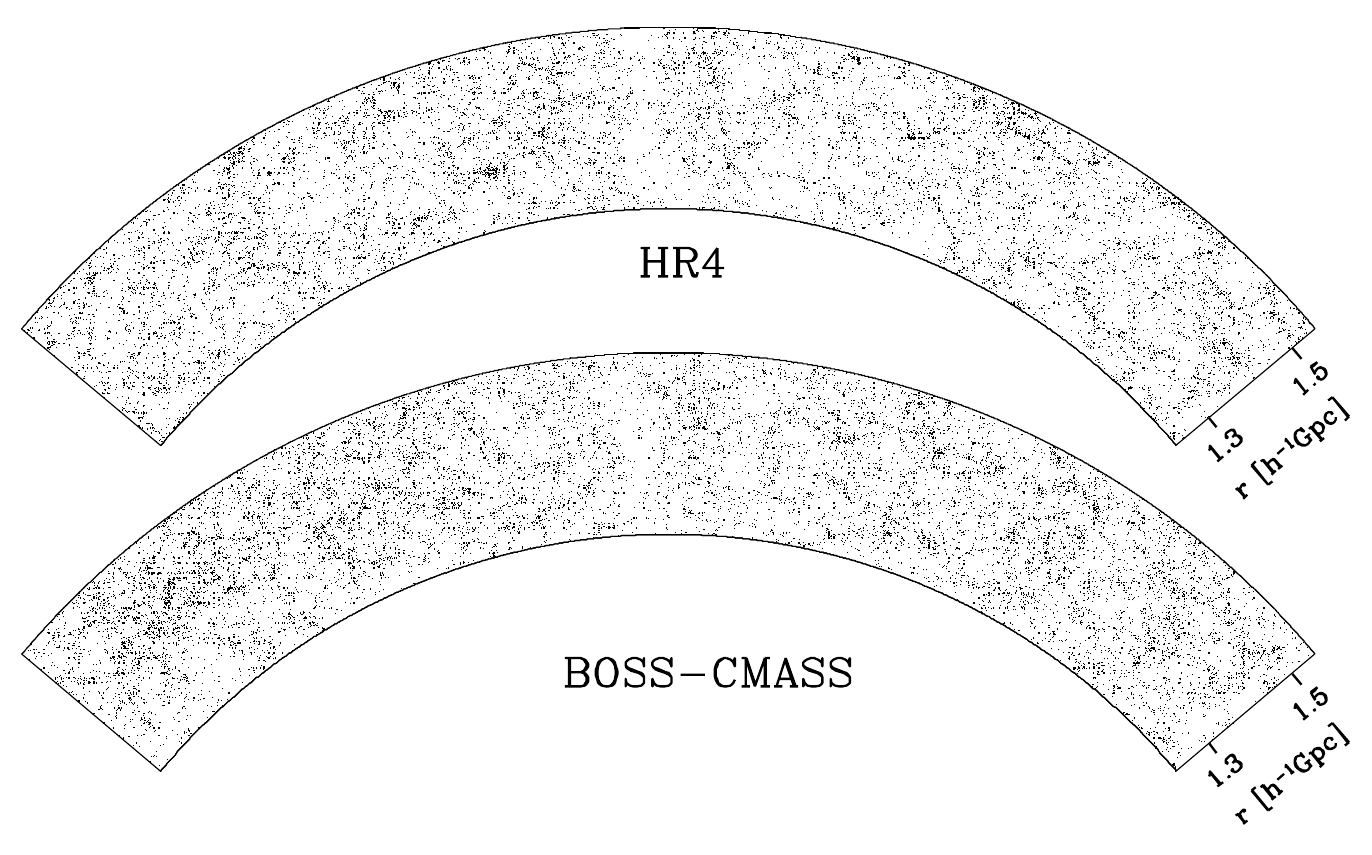}
\caption{
{\it Bottom}:  Distribution of galaxies with  $\mathcal{M}_r < -22.35$
and   $0.45<z<0.6$    in   the   BOSS-CMASS    volume-limited   sample
\citep{choi15}. 
Galaxies are  selected from a strip  with $-2^\circ \leq $  dec. $\leq
2^\circ$ and $130^\circ  \le $ R.A. $\le 230^\circ$ out  of the entire
BOSS survey area. 
{\it Top}: The  `galaxies' in the mock BOSS-CMASS  survey performed in
the HR4 simulation.  
The absolute magnitude of the observed galaxies is calculated with the
reference redshift of $z=0.55$ and is  used to produce the sample with
a constant number density.  
The HR4 PSB  subhalos are selected as galaxies in  accordance with the
galaxy-subhalo correspondence model, and  the galaxy number density at
a given  redshift is matched  with the  observed one by  adjusting the
low-mass cutoff.  
\label{figure} }
\end{figure*}

{To  generate  the  past  lightcone space  data,}  we  put  an
artificial observer at the origin $(x,y,z) =(0,0,0)$ of the simulation
box. 
At  each  time step,  we  calculate  the  comoving distance  from  the
observer using 
\begin{equation}
d_{\rm  c}  =  {c  \over   H_0}  \int_0^z  {1\over  E(z^\prime)}  {\rm
  d}z^\prime , 
\end{equation}
where 
\begin{multline}
  \label{eq:Ez}
  E(z) \equiv\\
  \sqrt{\Omega_{m,0}                     (1+z)^3                     +
    (1-\Omega_{m,0}-\Omega_{\Lambda,0})(1+z)^2 + \Omega_{\Lambda,0}}. 
\end{multline}
Then, we search for particles located in a comoving shell, whose inner
and  outer boundaries  at the  $i$-th step  are $d_{{\rm  c},i}-\Delta
d_{{\rm  c},i}/2$  and   $d_{{\rm  c},i}+\Delta  d_{{\rm  c},i+1}/2${,
  respectively,} where $\Delta d_{{\rm c},i+1} \equiv (d_{{\rm c},i+1}
- d_{{\rm c},i})/2$.  
We {utilize}  the periodic boundary conditions  by copying the
simulation box  to extend  the {all-sky} past  lightcone space
data up to $r=3150 ~h^{-1}{\rm Mpc}${, which corresponds to $z
  \simeq 1.5$}. 

Due  to  the finite  step  size,  several  undesirable events  may  be
encountered.  
If a particle crosses the  shell boundary between two neighboring time
steps, it  can be missed or  be counted twice in  the lightcone space
data. 
Therefore, we set a buffer zone laid upon both sides of the shell. 
The width of the buffer zone is  determined to be equal to the maximum
displacement taken by a particle in a time step. 
Using these buffer zones, we can catch these crossing events.  
A crossing particle  can simultaneously be detected  in two contacting
shells  or  two  adjacent  buffer  zones,  and  we  simply  merge  the
duplicated  particle  by  averaging   position  and  velocity  in  the
lightcone space data.

{In both the snapshot and past lightcone data,} we apply the
Ordinary  Parallel  {Friend-of-Friend}   (OPFOF),  a  parallel
version of {FoF} code to identify virialized halos. 
The standard percolation length is simply adopted as $l_{\rm link}=0.2
d_{\rm mean}$. 
The  halo position  and peculiar  velocity  are given  as the  average
position and velocity of the member particles. 
Then    we   apply    the   {physically    self-bound   (PSB)}
{sub}halo finding method \citep{kim and park 2006} to identify
subhalos embedded in the FoF halo. 
It  employs a  negative  total energy  criterion  and spherical  tidal
boundaries to discard particles from subhalo candidates.

As   a  representative   example,  Figure~\ref{figure}   compares  the
volume-limited galaxy  sample from BOSS-CMASS with  $r$-band magnitude
limit  $\mathcal{M}_r  < -22.35$  at  $0.45  \leq  z \leq  0.6$  ({\it
  bottom}) with the  mock galaxy sample from the HR4  built by the PSB
subhalo-galaxy correspondence model \citep[{\it top};][]{kim08}.

\subsection{Halo Merger Data}

To  build  the  merger  trees  we detect  halos  and  subhalos  at  75
equally-spaced sparse time steps from $z = 12$ to 0.  
The  step size  is  set  to be  comparable  to  the rotational  period
(i.e. dynamical timescale) of Milky-Way-size galaxies. 
Halo merger trees  are then built by tracing  the gravitationally most
bound member particles (MBPs) of halos. 
If a halo does  not contain any former MBP, we select  a new MBP among
the member particles of the halo.  
If  one   MBP  is  found,   we  assume  the   halo  to  be   a  direct
{descendant} of the halo. 
If the halo hosts multiple former MBPs,  we treat the halo as a merger
remnant and those  ancestor MBPs (or halos) are linked  to the remnant
creating a halo merger tree. 
These merger trees will be extensively used to build mock galaxies and
to compare with observations \citep{hong15}. 
Of course, due to the halo  mass resolution of the HR4 ($M_s=2.7\times
10^{11} ~h^{-1}{\rm  M_\odot}$), we are  unable to resolve  mergers of
sub-Milky-Way-mass (sub)halos.

\section{Properties of FoF Halos}

\subsection{Multiplicity Function}

The multiplicity function is defined as 
\begin{equation}
f(\sigma,z) \equiv {M\over \rho_b(z)} {{\rm d} n(M,z) \over {\rm d}\ln
  \left[ 1/\sigma(M,z) \right]} , 
\end{equation}
where  $n(M,z)$  is   the  cumulative  halo  mass   function  at  $z$,
$\rho_b${$(z)$}   is  the   background  matter   density,  and
{$\sigma(M,z)$}  is the  density fluctuation  measured on  the
mass scale of $M$.  
For  a  given  power  spectrum  $P(k)$,  the  density  fluctuation  is
estimated as 
\begin{equation}
\sigma^2(M,z) =  {D_1^2(z)\over 2\pi^2}  \int k^2  P(k) |W(kR(M,z))|^2
{\rm d}k, 
\end{equation}
where 
\begin{equation}
R(M,z) \equiv \left( { 3 M \over 4\pi \rho_b(z)}\right)^{1/3}, 
\end{equation}
and
$D_1(z)$ is the  growing mode {of} the linear  growth factors computed
as 
\begin{equation}
D_1(z) = {5\over2}\Omega_{m,0}  E(z)\int_{z}^\infty {(1+z^\prime) {\rm
    d}z^\prime\over E^3(z^\prime)}. 
\end{equation}

\begin{table*}[tp]
\centering
\caption{Description   of   fitting   models   of   the   multiplicity
  function \label{tabmultiplicity}} 
\begin{tabular}{lll}
\toprule
Model & $f(\sigma, z)$ & Parameters \\
\midrule 
\cite{sheth99}$^{*}$ & 
$\displaystyle A \sqrt{{2 \over \pi}} \chi \left( 1+\chi^{-2p} \right)
\exp\left[-{\chi^2 \over 2}\right]$ & 
$(A, p, q) = (0.3222, 0.3, 0.707, 0.3)$ \\ 
\citet{jenkins01} & 
$\displaystyle A \exp\left(-\abs{\ln \sigma^{-1}+a}^b\right)$ & 
$(A,a,b) = (0.315,0.61,3.8)$\\ 
\cite{warren06} &
$\displaystyle     A    \left(     {\sigma^{-a}    +     b    }\right)
\exp{\left[-{c\over\sigma^2}\right]}$ & 
$(A, a, b, c) = (0.7234, 1.625, 0.2538, 1.1982)$ \\ 
\cite{tinker08}$^{\dagger}$ & 
$\displaystyle A  \left( {\sigma^{-a} + b}  \right) \exp\left[-{c\over
    \sigma^2}\right]$ & 
$(A, a, b, c) = (0.745, 1.47, 0.250, 1.19)$ \\
\cite{crocce10}$^{\dagger}$ & 
$\displaystyle A \left( {\sigma^{-a} + b }\right)
\exp{\left[-{c\over\sigma^2}\right]}$ &
$(A, a, b, c) = (0.58, 1.37, 0.30, 1.036)$ \\
\cite{manera10}$^{* \dagger}$ &
$\displaystyle A \sqrt{{2 \over \pi}} \chi \left( 1+\chi^{-2p} \right)
\exp\left[-{\chi^2 \over 2}\right]$ & 
$(A, p, q) = (0.3222, 0.248, 0.709)$ \\
\cite{bhattacharya11}$^*$ &
$\displaystyle     A\sqrt{2\over     \pi}    \chi^r     (1+\chi^{-2p})
\exp\left[-{\chi^2 \over2}\right]$ & 
$(A, p, q, r) = (0.333, 0.807, 0.788, 1.795)$ \\
\cite{angulo12} &
$\displaystyle  A\left({B\over\sigma}  +1\right)^q  \exp\left[-{C\over
    \sigma^2}\right]$ & 
$(A, B, C, q) = (0.201, 2.08, 1.172, 1.7)$ \\
\cite{watson13} &
$\displaystyle A \left({\sigma^{-a} +  b } \right) \exp\left[-{c \over
    \sigma^2}\right]$ & 
$(A, a, b, c) = (0.589, 2.163, 0.479, 1.210)$ \\
\bottomrule
\end{tabular}
\tabnote{
$^{*}$  $\chi   \equiv  \sqrt{q}  \delta_{\rm  c}   /  \sigma$,  where
$\delta_{\rm c} = 1.686$ is the density contrast at the collapse epoch
{in an Einstein-de Sitter universe}. \\ 
$^{\dagger}$  Only the case at $z = 0$ is given here.}
\end{table*}

\begin{figure}[tp]
\centering
\includegraphics[width=8.4cm]{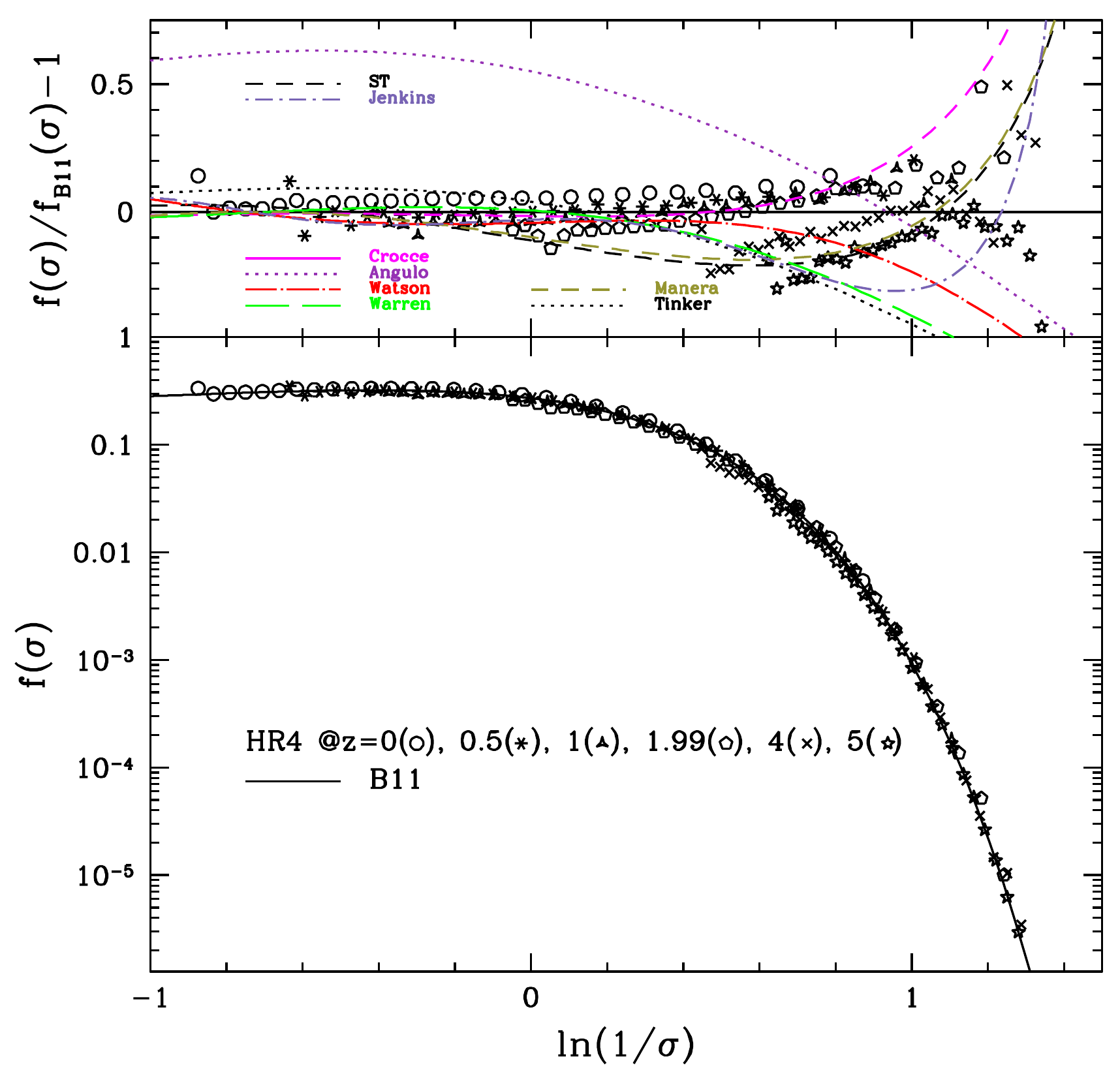} 
\caption{%
{\it Bottom}: Multiplicity functions {of HR4 FoF halos} at $z=0$, 0.5,
1, 1.99, 4, and 5.  
The solid curve is proposed by \citet{bhattacharya11}, B11.  
{\it Top}: Fractional deviations of  the simulated ({\it symbols}) and
modeled ({\it lines}) multiplicity functions from B11. 
\label{fofresc}
}
\end{figure}

Figure~\ref{fofresc} shows  the multiplicity function from  the HR4 as
well as a number of previous fitting models of the multiplicity function
(see Table~\ref{tabmultiplicity} and references therein). 
Top panel shows the deviations  of multiplicity functions with respect
to the model described in \citet{bhattacharya11}, hereafter B11.  
For clarity,  in the cases  of \cite{crocce10} and  \cite{manera10} we
only show the fitting function obtained at $z=0$. 
All  the previous  fitting functions  significantly deviate  from each
other at high mass scales. 
This may be produced by the exponential cut off producing large noises
in fitting the steep {slope}.  
From  the simulated  multiplicity  function, we  can  clearly see  the
redshift change, and therefore  a single functional  form may  not be
sufficient. 
For  large  values of  $\ln  (1/\sigma)$,  the redshift  evolution  of
multiplicity functions is substantial  \citep{lukic07}, and the overall
amplitude seems to increase as the redshift decreases.

\begin{figure}[tp]
\centering
\includegraphics[width=8.4cm]{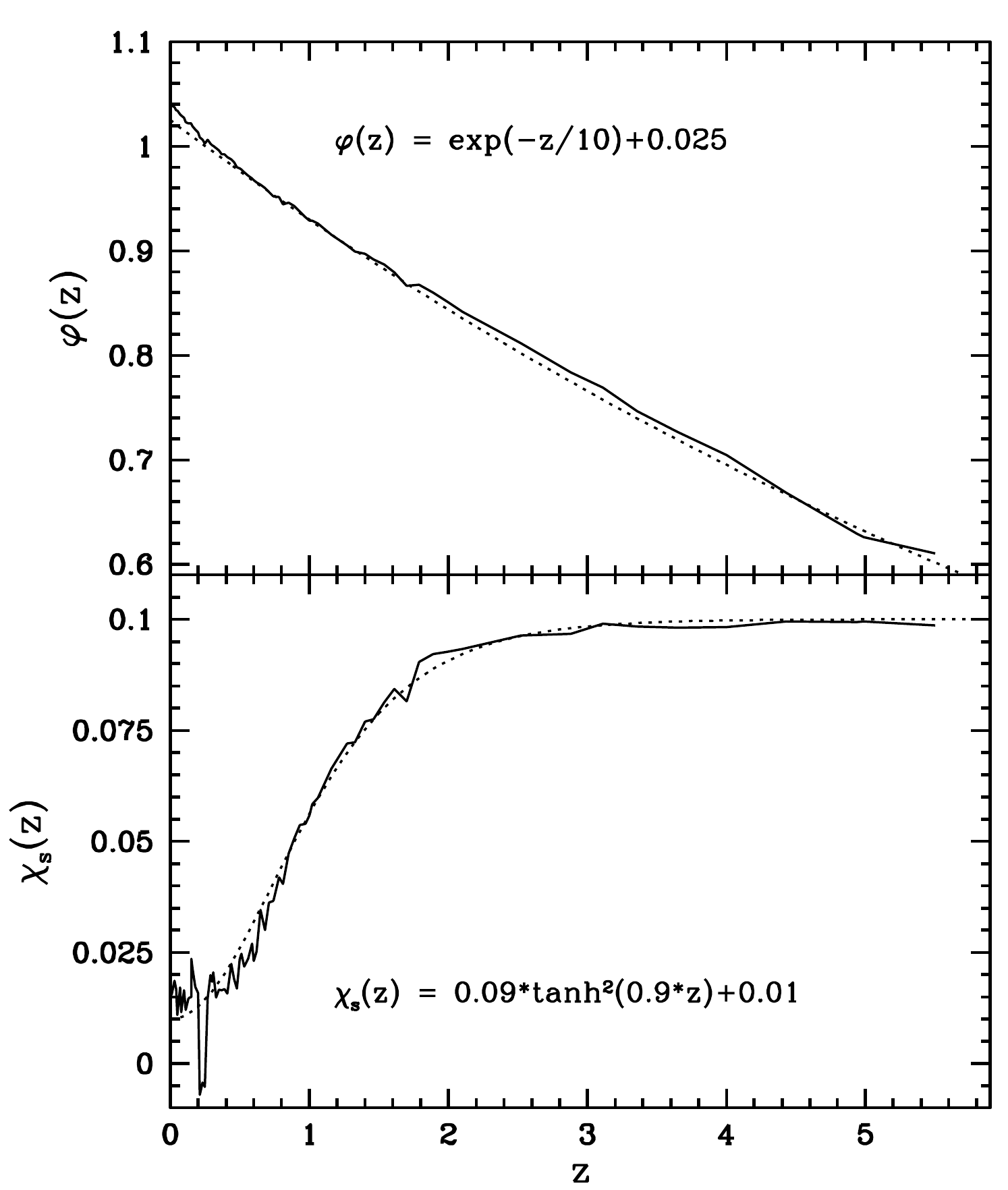} 
\caption{
Redshift-dependent  $\chi$-  and   amplitude-corrections  in  Equation
\ref{kimeq}  showing  the  best  fit   to  the  HR4  FoF  multiplicity
functions. 
The  dotted lines  are  the  analytic fitting  functions  as shown  in
Equations (\ref{chiz}) and (\ref{psiz}). 
\label{exefit}
}
\end{figure}

We fit the  simulated multiplicity function with a variant  of the B11
function with an amplitude changing with redshift as 
\begin{eqnarray}
f_{\rm  Kim}(\chi_{\rm  L},z) \equiv  \varphi(z)f_{\rm  B11}(\chi_{\rm
  L}(M,z) -\chi_s(z)). 
\label{kimeq}
\end{eqnarray}
Here   {$\chi_{\rm   L}(M,z)   \equiv  \sqrt{q}   \delta_{\rm   c}   /
  \sigma(M,z)$},  where  $\delta_{\rm  c}  =  1.686$  is  the  density
contrast at the collapse epoch  in an Einstein-de Sitter universe, $q$
is    a    fitting    parameter    in   the    B11    function    (see
Table~\ref{tabmultiplicity}),  and  $\chi_s(z)$ and  $\varphi(z)$  are
redshift-dependent $\chi$- and amplitude-corrections, respectively.  
The value  of $\delta_\text{c}$ in  an Einstein-de Sitter  Universe is
1.686, and slightly depends on the cosmology.  
However, for consistency with previous work, we will use this constant
value of 1.686 \citep[e.g.,][]{bhattacharya11}.  
We  fit  the simulated  multiplicity  function  with the  least-square
minimization and obtain the empirical fitting function as 
\begin{eqnarray}
\label{chiz}
\chi_s(z) &=& 0.09 \tanh^2( 0.9z ) + 0.01 \\
\label{psiz}
\varphi(z) &=& \exp\left(-{z\over 10}\right) + 0.025.
\end{eqnarray}
Figure \ref{exefit} shows the redshift evolution of $\chi_s(z)$ (left)
and $\varphi(z)$, respectively. 
At high redshifts, the HR4 simulation underpopulates halos compared to
B11, while the HR4 has more halos than B11 in the recent epoch. 
Also, as we move to higher redshift, $\chi_s (z)$ increases and reaches
about 0.098 at very high redshift.

\begin{figure}[tp]
\centering
\includegraphics[width=8.4cm]{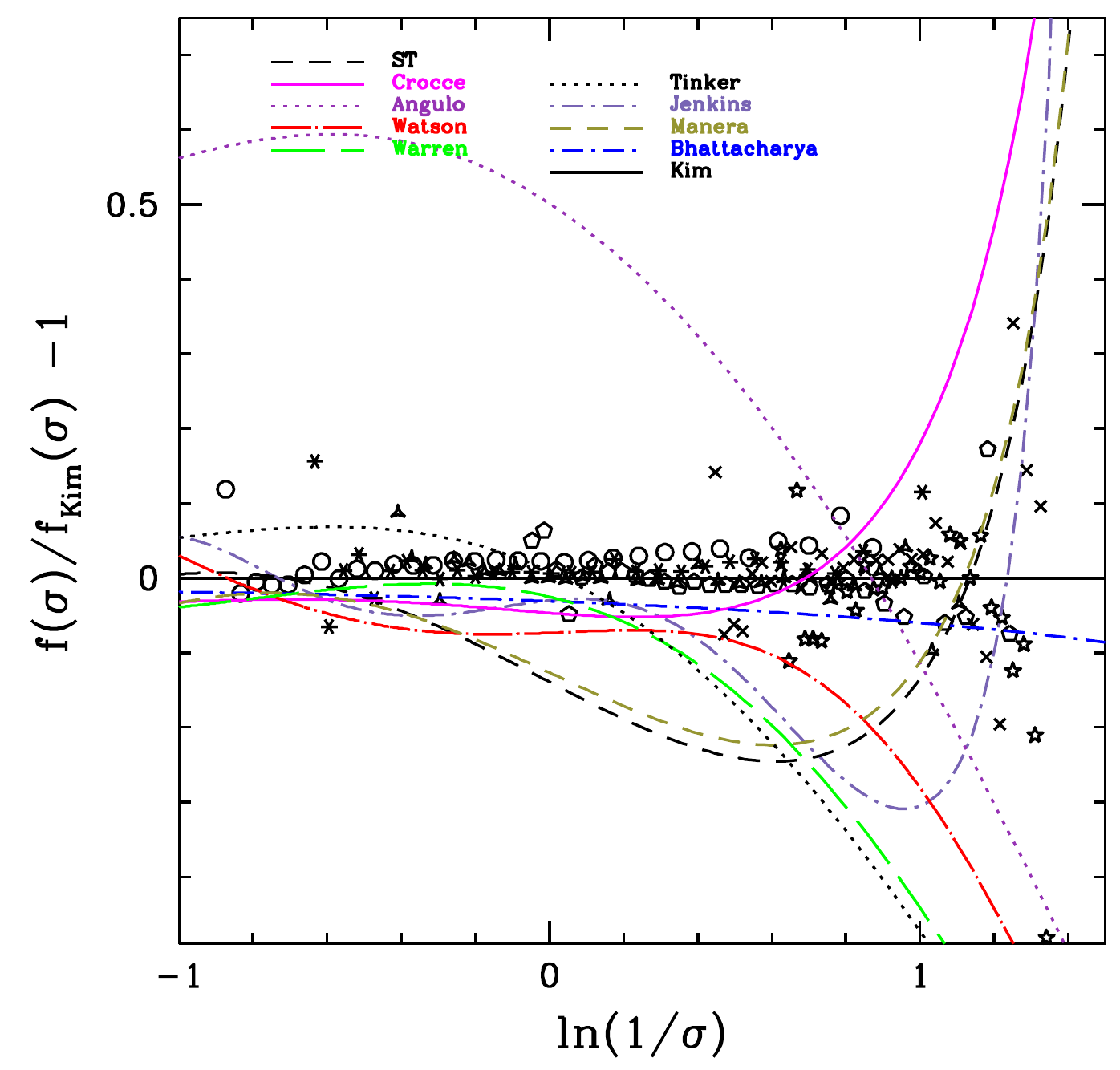} 
\caption{
Simulated and modeled multiplicity functions  with respect to {our new
  fitting model ($f_{\rm Kim}(\sigma, z)$).}  
Same symbol and color conventions as in Figure \ref{fofresc}.
\label{fofrescfinal}
}
\end{figure}

Figure~\ref{fofrescfinal}   shows  the   scatter  of   the  simulated
multiplicity function  ({symbols}) over  {our} fitting
model  ($f_{\rm   Kim}$)  with  various  former   models  (lines)  for
comparison.  
While other  fitting models match the  simulated multiplicity function
only at small  scales ($\ln (1/\sigma) < 0.3$), our  new fitting model
agrees with the HR4 on most scales  in the redshift range between $z =
5$ and 0, within a 2.5\% fluctuation level.  

\begin{figure}[tp]
\centering
\includegraphics[width=8.4cm]{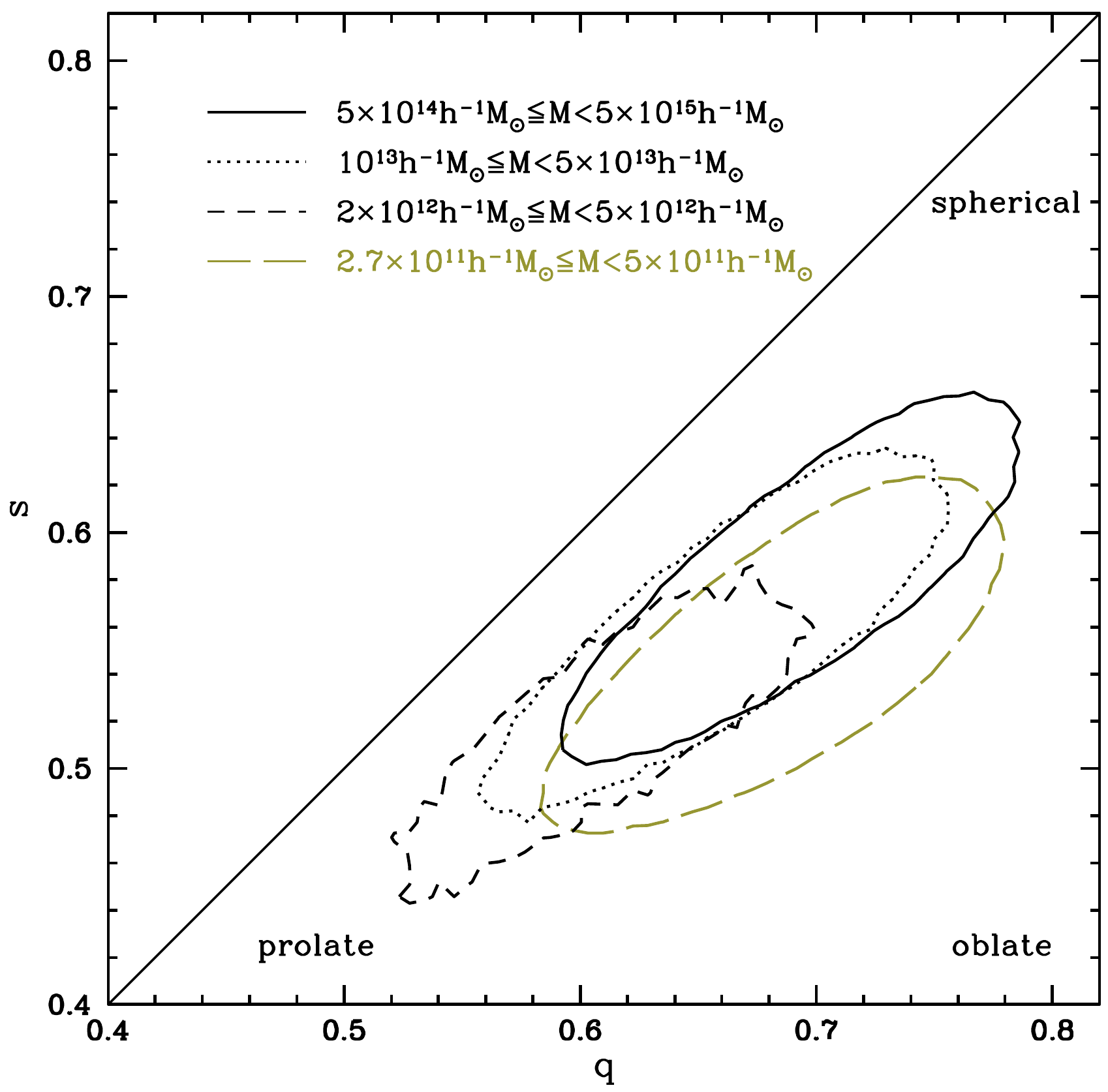} 
\caption{Shape  distributions of  FoF  halos in  various mass  samples
  shown in the $q$-$s$ diagram at $z=0$. 
We  mark  iso-density  contours   enclosing  25\%  halos  around  peak
distributions. 
\label{qs}
}
\end{figure}

The origin of  the redshift evolution of the  multiplicity function is
not clearly  known but it might  be partly explained by  the following
arguments. 
First, even the 2LPT may be somewhat insufficient to generate accurate
initial conditions of the simulation.  
Such   effect   would   be   avoided   by   introducing   higher-order
approximations  (for   comparison  between   the  Zel{'}dovich
approximation and 2LPT, see \citealt{crocce06}). 
However, the  target redshifts to  measure the halo mass  function are
sufficiently  lower  ($z\lesssim  5$;  see  the  discussions  made  by
\citealt{tatekawa07}) compared to the  initial epoch of $z_i=100$ and,
consequently, the  redshift evolution may  not be caused  by numerical
transients.  
Second, it  may be due  to the  limitation of the  linear perturbation
theory    or    the    linear     growth    model    in    calculating
{$\sigma(M,z)$}  after  the {nonlinear}  gravitational
clustering begins to enter.  
The multiplicity  function {assumes} that  there is no  other redshift
dependence than the matter fluctuation, $\sigma(M,z) = D(z)\sigma(M)$.  
But as the non{-}linear growth becomes significant at lower redshifts,
one should consider the effect of non{-}linear gravitational evolution
of density fields.

\subsection{ Halo Shape }
\subsubsection{Structure}

The shape tensor $S_{ij}$ of an FoF halo is defined
\begin{equation}
S_{ij} = \sum_{k}^{N_m} ({x^k_i - {\bar x}_i })( {x^k_j -{\bar x}_j}),
\end{equation}
where $i$ and  $j$ are structural axes, ${\bar x}$  is the position of
the center of the halo mass and $N_m$ is the number of member particles. 
The three eigenvalues of the shape  tensor $r_3 \leq r_2 \leq r_1$ are
respectively  the {lengths  of the}  minor, intermediate,  and
major axes of the corresponding ellipsoid. 
The prolateness and sphericity of a halo are defined as
\begin{eqnarray}
q & \equiv & { r_2 \over r_1 } \\
s & \equiv & { r_3 \over r_1 } .
\end{eqnarray} 
A halo is respectively defined as prolate, oblate, or spherical if
\begin{eqnarray}
&& q \ll 1, \\
&& q \simeq 1, s \ll 1, \\
&& s \simeq 1 .
\end{eqnarray}

Figure~\ref{qs} shows the probability distributions of ($q$, $s$) with
a contour  containing 25\%  of halos around  the peak  distribution in
four different mass samples of FoF halos.  
For  halo samples  more  massive than  $2  \times 10^{12}~h^{-1}  {\rm
  M_\odot}$, we can clearly see that  halos become more prolate as the
mass   increases,   in    agreement   with   theoretical   predictions
\citep[e.g.][]{rossi11}.  
Less massive halos with $M_s \leq M < 5 \times 10^{11}~h^{-1} {\rm
  M_\odot}$  have  their  distribution substantially  shifted  to  the
lower-right corner  in this diagram,  i.e., are more oblate  than more
massive samples.  
This may be fully explained by the particle discreteness effect, which
will be described {in the next section}.  

\begin{figure}[tp]
\centering
\includegraphics[width=8.4cm]{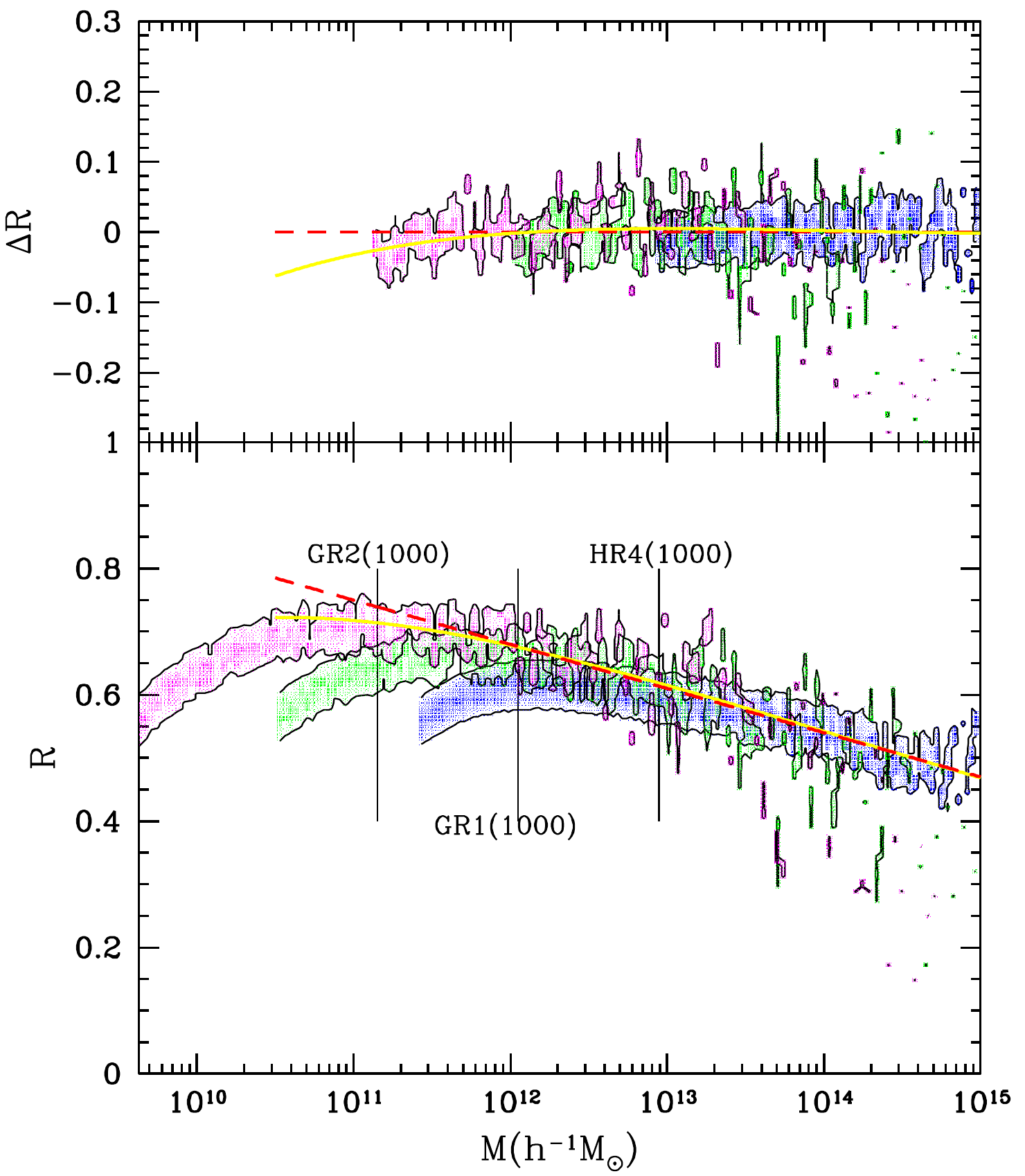} 
\caption{Resolution dependence of the roundness parameter at $z=0$
for HR4 ({\it blue}), GR1 ({\it green}), and GR2 ({\it magenta})
containing 25\% of halos around peak probabilities. 
{\it Bottom}: {Roundness parameter as a function of halo mass.}
A vertical bar marks the mass scale equivalent to 
the mass of 1000 particles {($10^3 m_{\rm p}$)} combined.
{The  best-fitting functions  of  $\mathcal{R}(M)$ from  halos
  with  lower-mass cutoff  $10^3 m_{\rm  p}$ ($\mathcal{R}_{1000}^{\rm
    fit}$;   {\it    yellow})   and    $5\times   10^3    m_{\rm   p}$
  ($\mathcal{R}_{5000}^{\rm fit}$; {\it red dash}) are shown.} 
{\it Top}: Deviation of the roundness parameter from $\mathcal{R}^{\rm
  fit}_{5000}$ with respect to the halo mass. 
Here we do not show the  distribution below the mass of {$10^3
  m_{\rm p}$}. 
\label{comsphericity} 
}
\end{figure}

\subsubsection{Roundness}
We now investigate the FoF halo shape from a different angle. 
First, we define the roundness as 
\begin{equation}
\mathcal{R} \equiv \sqrt{qs} = \sqrt{{r_2 r_3 \over r_1^2}} .
\end{equation} 
To measure the resolution effects on halo shape, we ran two additional
higher-resolution simulations called Galaxy Run 1 (GR1) and Galaxy Run
2 (GR2). 
These simulations used $2048^3$ particles.  
We employed the  same cosmological model but  different simulation box
sizes  ($L_{\rm box}^{\rm  GR1}  = 512{~h^{-1}{\rm  Mpc}}$ \&  $L_{\rm
  box}^{\rm GR2} = 256 ~h^{-1}{\rm Mpc}$). 
The mean particle  separations of GR1 and GR2  are $d_{\rm mean}=0.25$
and $0.125~h^{-1}{\rm Mpc}$, 
respectively, while the corresponding force resolutions are changed to
$0.025$ and $0.0125 {~h^{-1}{\rm Mpc}}$ accordingly. 
Therefore,  the mass  and force  resolutions are  quite enhanced  with
respect to the HR4. 

Figure~\ref{comsphericity} shows the  distribution of $\mathcal{R}$ of
FoF halos at $z=0$. 
In each simulation, at scales of $M \gg 10^3 m_{\rm p}$,
$\mathcal{R}$ tends to be independent of the simulation resolution.  
On the other hand, at small  mass scales ($M \lesssim 10^3 m_{\rm p}$)
each simulation seems to  underestimate $\mathcal{R}$, probably due to
the small number of particles. 
\cite{hoffmann14} examined  the discreteness effect on  a modeled halo
for a given shape 
and found  that the required  number of  particles should not  be less
than 1000  for a reliable  shape determination  (see Fig. A2  in their
paper).

From our three simulations we found a fitting formula of the roundness
parameter as  a function of  the halo mass  for massive halos  with $M
\geq 10^3 m_{\rm p}$: 
\begin{multline}
{\mathcal{R}}^{\rm fit}_{1000} (M) = a_\mathcal{R} \log_{10} \left({ M
    \over 10^7~h^{-1}{\rm M_\odot} }\right) \\ 
\times   \exp  \left[   -b_\mathcal{R}  \log_{10}   \left({  M   \over
      10^{7}~h^{-1}{\rm M_\odot} }\right) \right] , 
\label{R1000}
\end{multline}
where $(a_{\mathcal{R}}, b_{\mathcal{R}}) = (0.55, 0.28)$ (yellow line
in Figure~\ref{comsphericity}). 
One  should  note  that  this  fitting is  only  valid  for  $M\gtrsim
1.4\times 10^{11} ~h^{-1}{\rm M_\odot}$, 
which is set by the combined mass of 1000 particles of the GR2.  
We do  not find  any turn-around  mass scale  of $\mathcal{R}$  in the
available mass range in this study. 
If we only consider  halos with $M \geq 5 \times  10^3 m_{\rm p}$, the
distribution of $\mathcal{R}$ follows 
\begin{equation}
\mathcal{R}^{\rm fit}_{5000} (M) = A_{\mathcal{R}} \log_{10} \left({ M
    \over 10^{12}~h^{-1}{\rm M_\odot} } \right) + B_{\mathcal{R}}, 
\label{R5000}
\end{equation}
where $(A_{\mathcal{R}}, B_{\mathcal{R}}) = (-0.07, 0.68)$ (red dashed
line in Figure~\ref{comsphericity}). 
Similar to the case of $\mathcal{R}^{\rm fit}_{1000}$, the above fitting
is only valid for $M \gtrsim 7 \times 10^{11}~h^{-1}{\rm M_\odot}$. 
Both $\mathcal{R}^{\rm fit}_{1000}$ and $\mathcal{R}^{\rm fit}_{5000}$
describe well the change of $\mathcal{R}$,  but they diverge below $M =
7  \times 10^{11}~h^{-1}{\rm  M_\odot}$, the  mass scale  of about  $5
\times 10^3 m_{\rm p}$ of GR2. 
Therefore, we may need a simulation with a higher mass resolution than
GR2 to see which fitting  formula describes the roundness parameter of
low-mass halos.

\begin{figure}[tp]
\centering
\includegraphics[width=8.4cm]{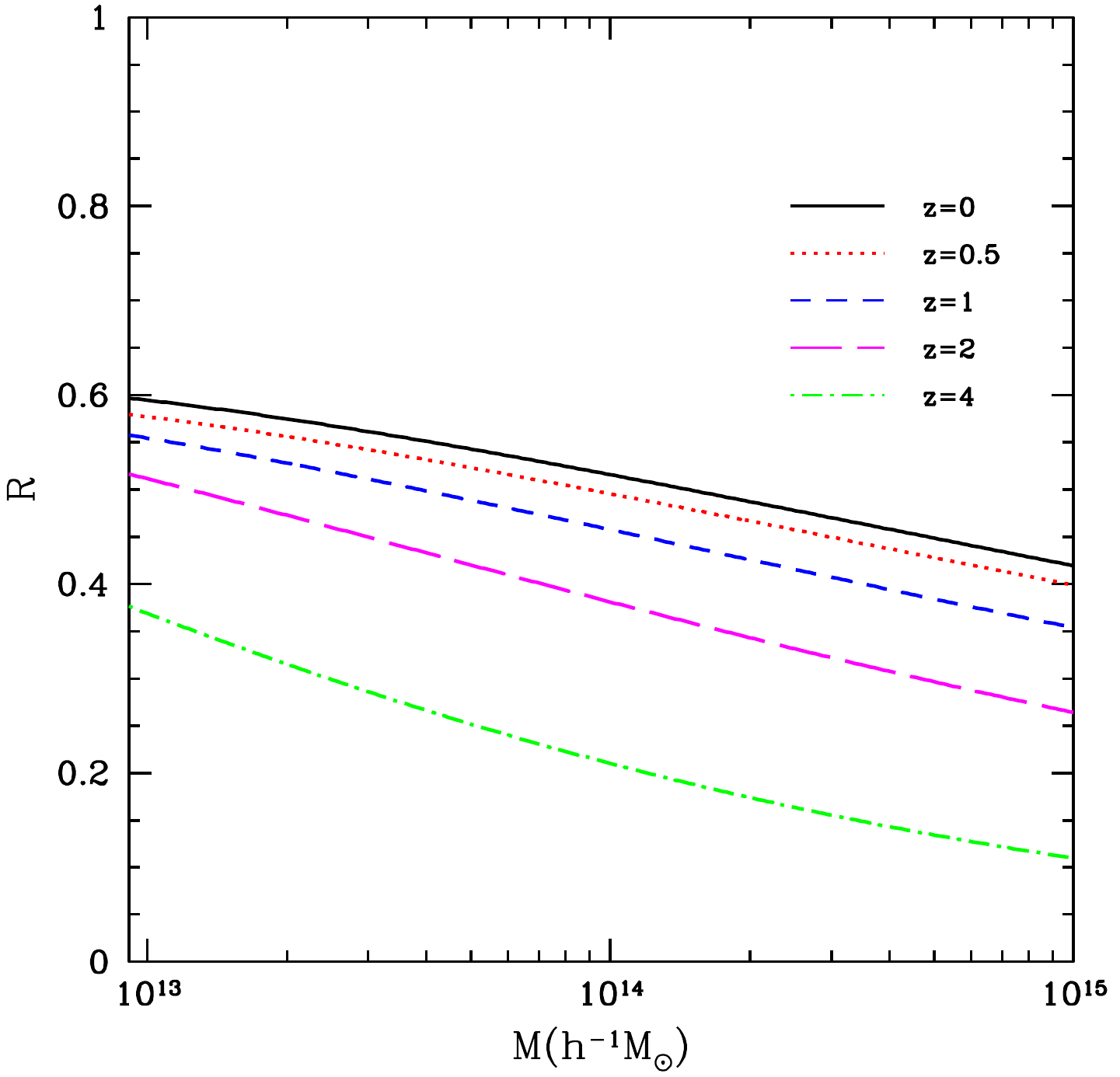} 
\caption{Change of peak position  in the $\mathcal{R}$ distribution at
  several redshifts in the HR4. 
The lower bound of the $x${-}axis corresponds to $10^3 m_{\rm p}$.
\label{sphericity_red}
}
\end{figure}
Figure~\ref{sphericity_red}   shows   the    redshift   evolution   of
$\mathcal{R}$ in the HR4. 
At higher redshift, halos tend to {have a smaller value of $\mathcal{R}$}
for a given mass.  
However, it is important to note that this tendency does not guarantee
a possible shape evolution of virialized halos because halos also grow
in mass with time.

\subsection{Halo Orientations}
In  this  section{,}  we  study the  angle  between  the  halo
rotational and structural axes. 
The directional angle between them is calculated as 
\begin{equation}
\theta_i       =      \cos^{-1}       |\hat{\boldsymbol{J}}      \cdot
\hat{\boldsymbol{r}}_i | , 
\end{equation}
where  $\hat{\boldsymbol J}$  is  the normalized  rotational axis  and
$\hat{\boldsymbol r}_i$  is the  unit vector in  the direction  of the
structural axis $i$. 
We  define the  probability distribution  function of  the directional
angles 
\begin{equation}
p^\prime(\theta) \equiv { {\rm d} P(\theta) \over {\rm d} \cos \theta} ,
\end{equation}
where $P(\theta)$ is the cumulative probability of a directional angle
greater than $\theta$. 
Then, for a random orientation  $p^\prime(\theta)$ is uniform over the
angle of $0^\circ \le \theta < 90^\circ$. 

\begin{figure}[tp]
\centering
\includegraphics[width=8.4cm]{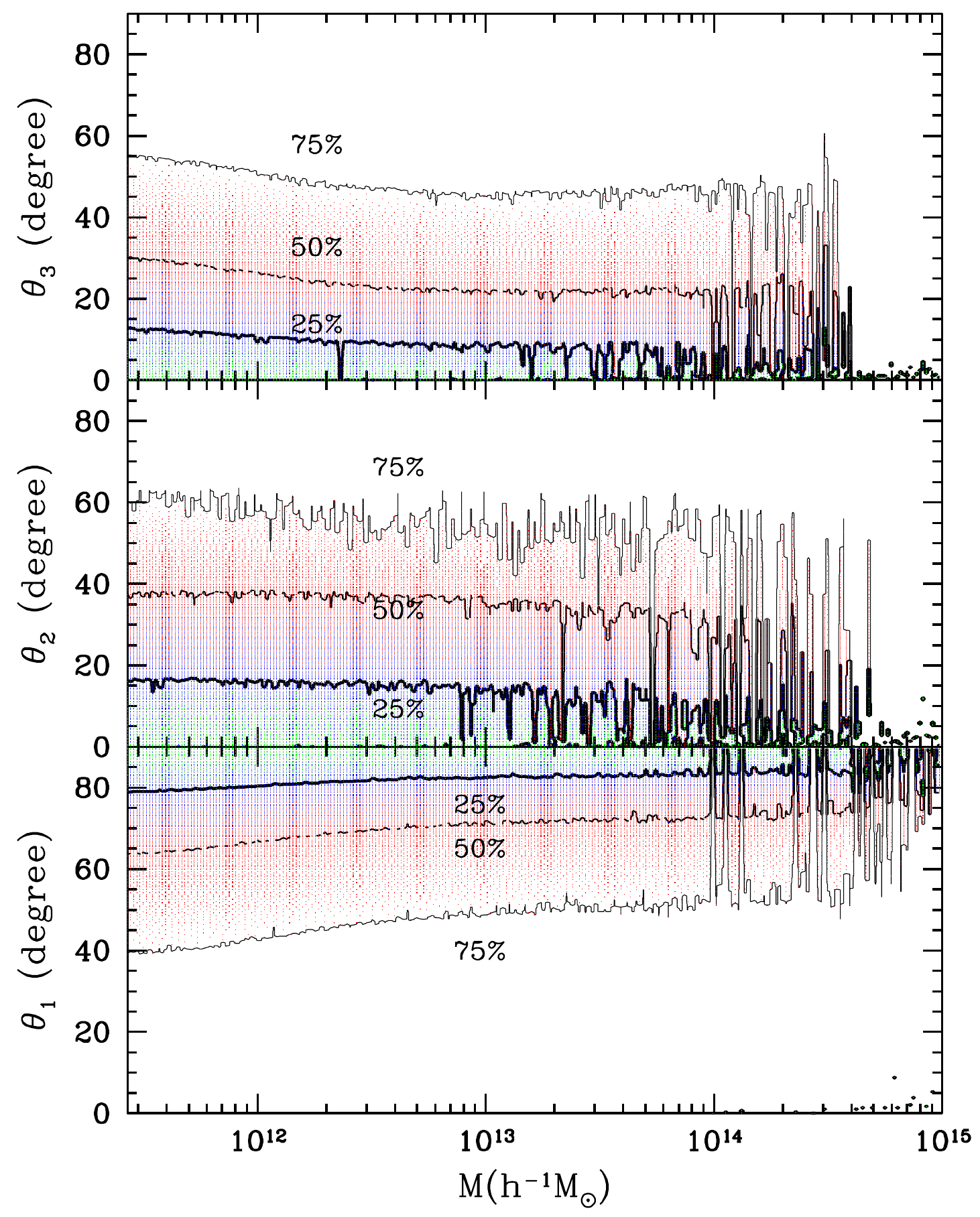} 
\caption{Orientations of {the} rotational axis with respect to
  the  major ({\it  bottom}), intermediate  ({\it middle}),  and minor
  ({\it top panel}) axes at $z=0$.  
Probability contours are  drawn around the peak position  at each mass
bin enclosing 25\%, 50\%, and 75\% of halos respectively.  
Most  halos are  positioned  at  $\theta_1\simeq 90^\circ$,  $\theta_2
\simeq 0^\circ$, and $\theta_3 \simeq 0^\circ$. 
\label{Orien}
}
\end{figure}

Figure~\ref{Orien} shows  the relations between the  rotation and halo
axes as a function of halo mass.  
The rotational axis  tends to be orthogonal to the  major axis (bottom
panel), which means that halos tend to swing around their major axis.  
Moreover,  from  the upper  two  panels,  it  can  be noted  that  the
rotational  axis is  more  aligned with  the minor  axis  than the
intermediate axis.  This alignment becomes stronger as the halo mass
increases. 
In addition, we find that this tendency still holds for low-mass halos
with ${M \lesssim} 10^3 m_{\rm p}$, implying that the halo rotation is less
affected by the mass resolution limit than the halo shape.

\section{Two-Point Correlation Function}

In this section, we implement the effects of redshift-space distortions
on the clustering of mock galaxies and measure the change of clustering in
the radial ($\pi$) and tangential ($\sigma$) directions. 
In the 3-dimensional  space, the radial separation  between two points
(${\boldsymbol r}_1$ and ${\boldsymbol r}_2$) is defined as
\begin{equation}
\pi   \equiv  \frac{\left|{\boldsymbol   d}_{12}  \cdot   {\boldsymbol
      R}_{12}\right|} { \left|{\boldsymbol R}_{12}\right|}, 
\end{equation}
where  ${\boldsymbol  R}_{12} \equiv  ({\boldsymbol  r}_1+{\boldsymbol
  r}_2)/2$  and  {${\boldsymbol  d}_{12} \equiv  {\boldsymbol  r}_1  -
  {\boldsymbol r}_2$. 
The tangential distance between them is simply obtained with 
\begin{equation}
\sigma   =\sqrt{  {\boldsymbol   d}_{12}\cdot{\boldsymbol  d}_{12}   -
  \pi^2}. 
\label{sigma}
\end{equation}
The correlation  function of a point  set can easily be  calculated by
Hamilton's method \citep{hamilton93}: 
\begin{equation}
\xi(\sigma,\pi)             =             {\boldsymbol{DD}(\sigma,\pi)
  {\boldsymbol{RR}}(\sigma,\pi) \over {\boldsymbol{DR}}(\sigma,\pi)^2}
- 1, 
\end{equation}
Here, 
$\boldsymbol{DD}$ is the number of pairs {of} real points, 
$\boldsymbol{DR}$ is  the number of  cross pairs between the  real and
random  points,   and  $\boldsymbol{RR}$   is  the  number   of  pairs
{of}  random  points  at  the  two-dimensional
separations of $\sigma$ and $\pi$.

We use the PSB subhalo catalog from the HR4 snapshot at $z = 0$ as our
mock galaxy sample. 
By  adopting  the  far-field  approximation  and  using  the  periodic
boundary   condition   of  the   HR4   simulation,   we  produce   the
redshift-space distortion in the $x$-direction, 
\begin{equation}
{ x}^\prime = { x} + {{ v_x}\over H_0}, 
\label{xdistort}
\end{equation}
where  $H_0$ is  the Hubble  parameter at  $z=0$ and  ${ v_x}$  is the
peculiar velocity along the $x${-}axis. 
 Since  we  adopt the  far-field  approximation,  $\pi$ is  the  position
 difference in the $x${-}axis and $\sigma$ is the 
 separation in the $y$-$z$ plane. 
We then construct a mass-limited  mock galaxy sample with PSB subhalos
satisfying $M \geq 2.60 \times 10^{12}~h^{-1}{\rm M_\odot}$. 
The average number  density of the mass-limited PSB  subhalo sample is
${\bar  n}   =  1.48  \times  10^{-3}~h^3{\rm   Mpc^{-3}}$,  which  is
comparable to} the number density  of the volume-limited sample of the
SDSS  Main galaxies  with absolute  magnitude limit  of $\mathcal{M}_r
-5\log_{10} h <-21$ \citep{choi10}. 

\begin{figure*}[tpb]
\centering
\subfigure[real space]{ \includegraphics[width=232pt]{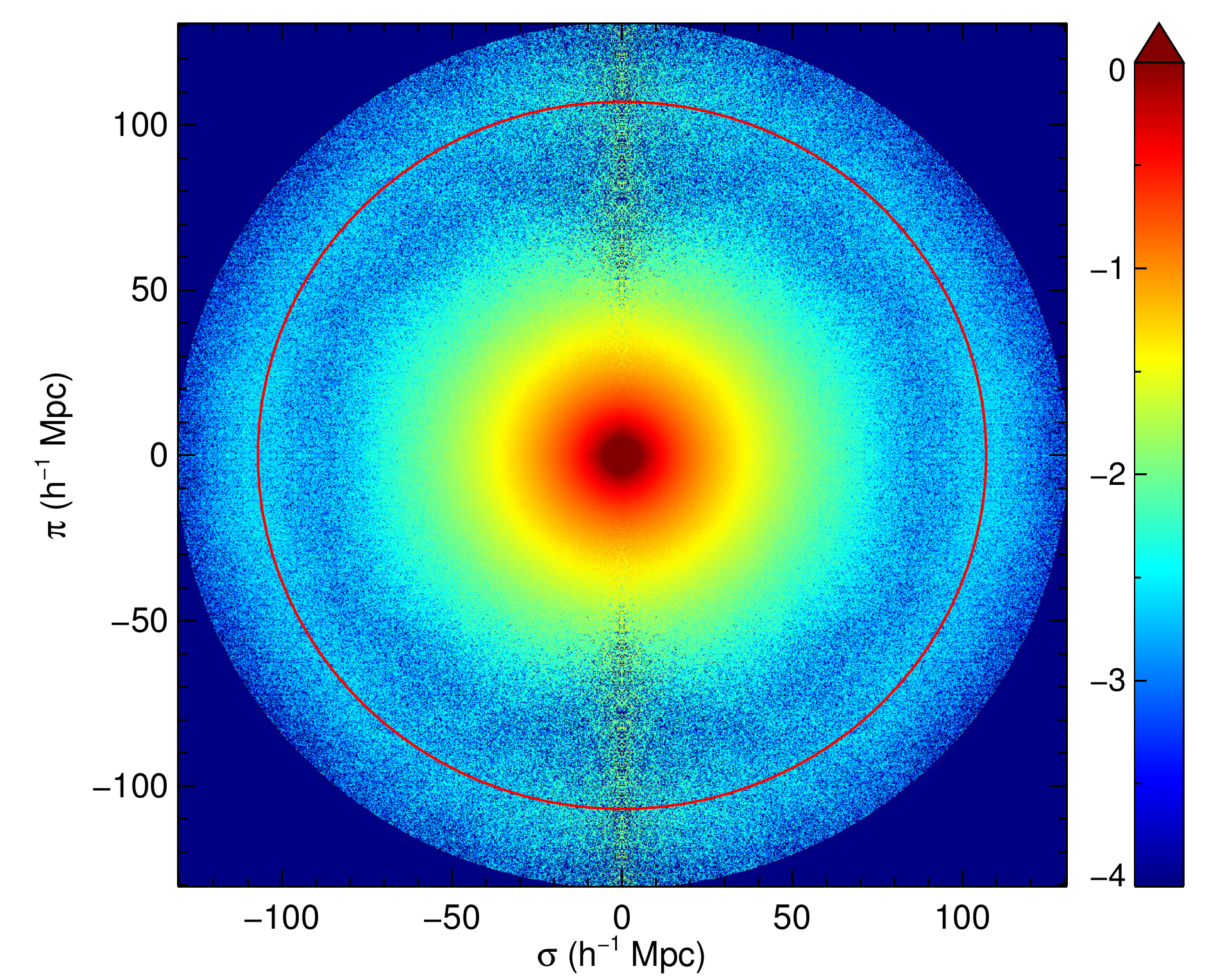} }
\subfigure[redshift space]{ \includegraphics[width=232pt]{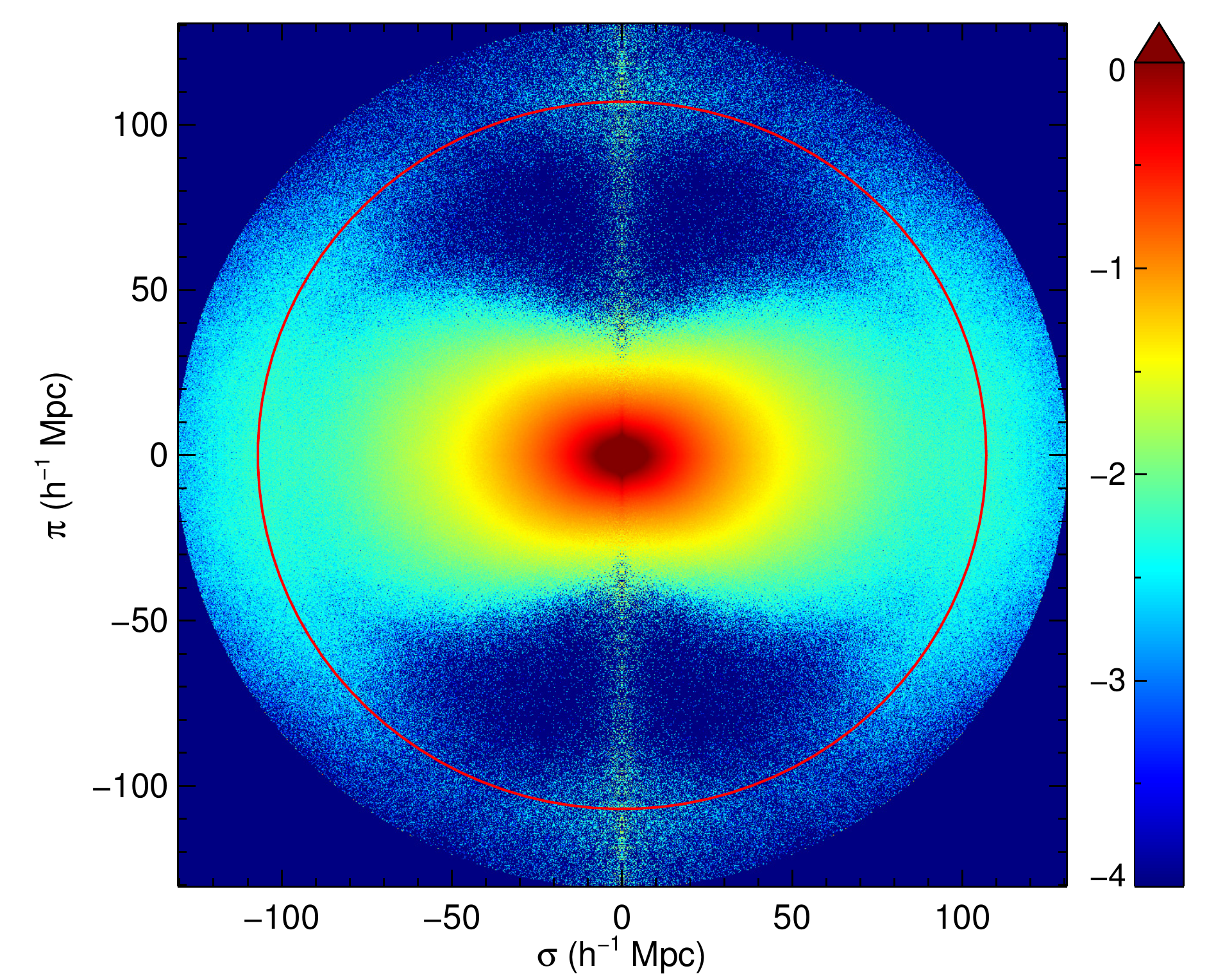} }
\caption{Correlation functions of mock galaxies measured without ({\it
    left}) and with ({\it right}) 
redshift-space distortion effects.
The   radius   of   each   circular   region   is   $130   ~h^{-1}{\rm
  Mpc}${,} and the solid circle marks 
the BAO peak position ($r_{\rm peak} \simeq 107 ~h^{-1}{\rm Mpc}$).
The color bar marks the correlation in logarithmic spacing.
\label{corr}
}
\end{figure*}

\begin{figure}[pb]
\centering
\includegraphics[width=8.4cm]{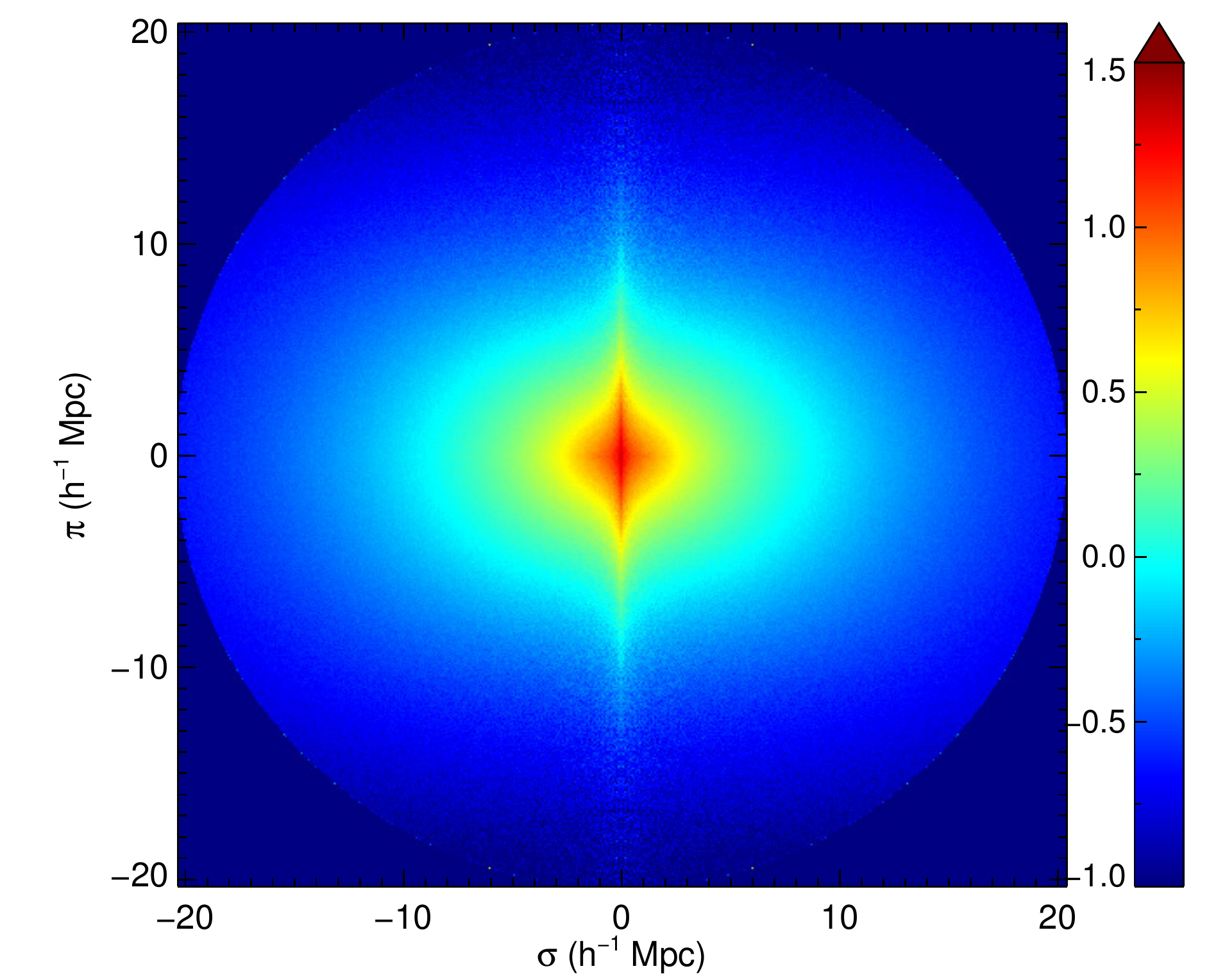} 
\caption{%
Same as the right panel of Figure~\ref{corr}, but zoomed in to clearly
show the finger-of-god effect. 
\label{corr_zoom}
}
\end{figure}

Figure~\ref{corr} shows  the effects  of redshift-space  distortions on
the correlation map. 
The left panel shows the correlation of our mock galaxy sample in real
space while  the effects of  redshift-space distortion are  applied in
the right panel. 
The   shape    of   $\xi(\pi,\sigma)$    is   distorted    along   the
line{-}of{-}sight    ({LoS,}    or   in    the
$\pi${-}direction). 
At the  very center, the  finger-of-god effect  can be seen  as spikes
stretching  along the  $\pi${-}direction  (for  a better  view
around the center, see Figure~\ref{corr_zoom}). 
On the  other hand, on  larger scales, the correlation  function along
the {LoS} contracts to the smaller scale.  

The position of the  BAO peak in real space can  be estimated from the
linear correlation function 
\begin{equation}
\xi_{\rm  linear}(r) \equiv  {1\over 2\pi^2}  \int k^2  P(k) {\sin(kr)
  \over kr} {\rm d}k. 
\end{equation}
For the WMAP  5-year standard $\Lambda$CDM cosmology,  the BAO
peak in real  space is located at $r_{\rm  peak} \simeq 107~h^{-1}{\rm
  Mpc}$,  shown as a solid circle in Figure~\ref{corr}.

\begin{figure}[tp]
  \centering
  \includegraphics[width=8.4cm]{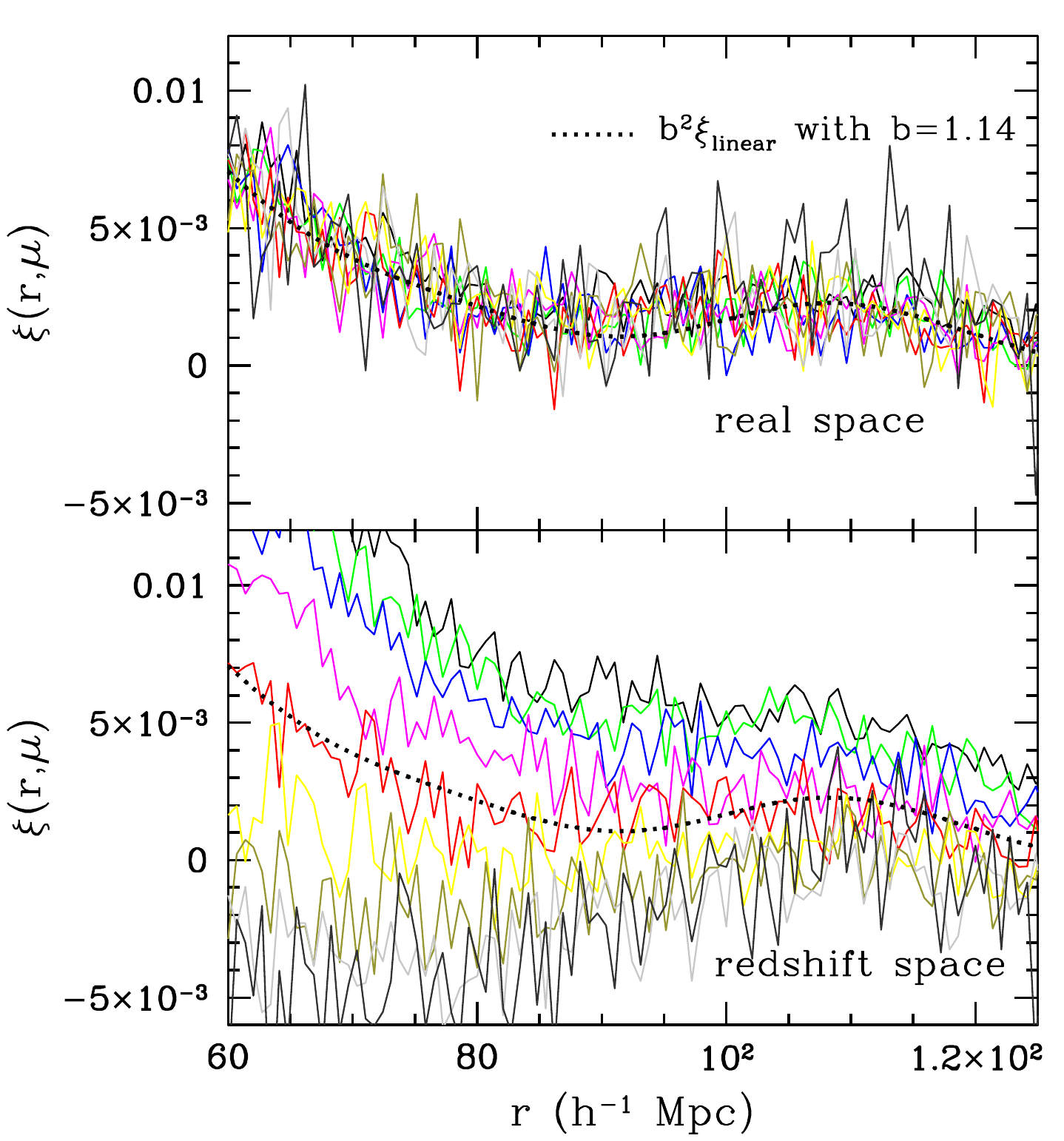} 
  \caption{Correlation functions between PSB  halo pairs separated along
    the directions of 
    $\theta=10$, 20, 30, 40, 50, 60, 70,  80, and 90 degrees in real space
    ({\it top}) 
    and redshift-distorted space ({\it bottom panel}).  
    In  the  bottom  panel,  the  top-most  line  is  the  correlation  of
    $\theta=80^\circ$, and  the correlation function increases  as $\theta$
    decreases. 
    The  dotted  line  is  the  linear  prediction  with  a  bias  factor
    $b=1.14$. 
    As the direction cosine ($\mu \equiv \cos\theta$) increases, the noise
    of the correlation functions is decreasing because the number of pairs
    along the given direction increases ($N_{\rm pair} \propto \mu$ ). 
    The  correlation functions  along $\theta=90^\circ$  {are}
    not shown due to big noise.  
    \label{bao}
  }
\end{figure}

Figure~\ref{bao}  shows   the  two-point  correlation   functions  for
different  values  of  the  directional cosine  to  the  {LoS}
direction $\mu$  in real  space ({\it top})  and redshift  space ({\it
  bottom panel}).  
In  real  space, the  correlation  function  around  the BAO  peak  is
independent  of  the  directional  angle  ($\theta$)  because  of  the
isotropic distribution.  
On  the other  hand, the  correlation functions  measured in  redshift
space are  increased as $\theta$  increases, because galaxy  pairs are
stretched along the {LoS}.  
It is  worth to  note that  the BAO peak  in the  tangential direction
($\theta = 90^\circ$) cannot be detected.  
Moreover, the correlation functions along the {LoS} has a peak
with   a  height   nearly   zero  while   correlation  functions   for
$\theta<30^\circ$  are  less  than  zero  on  scales  below  the  peak
position.

\begin{figure}[tp]
  \centering
  \includegraphics[width=8.4cm]{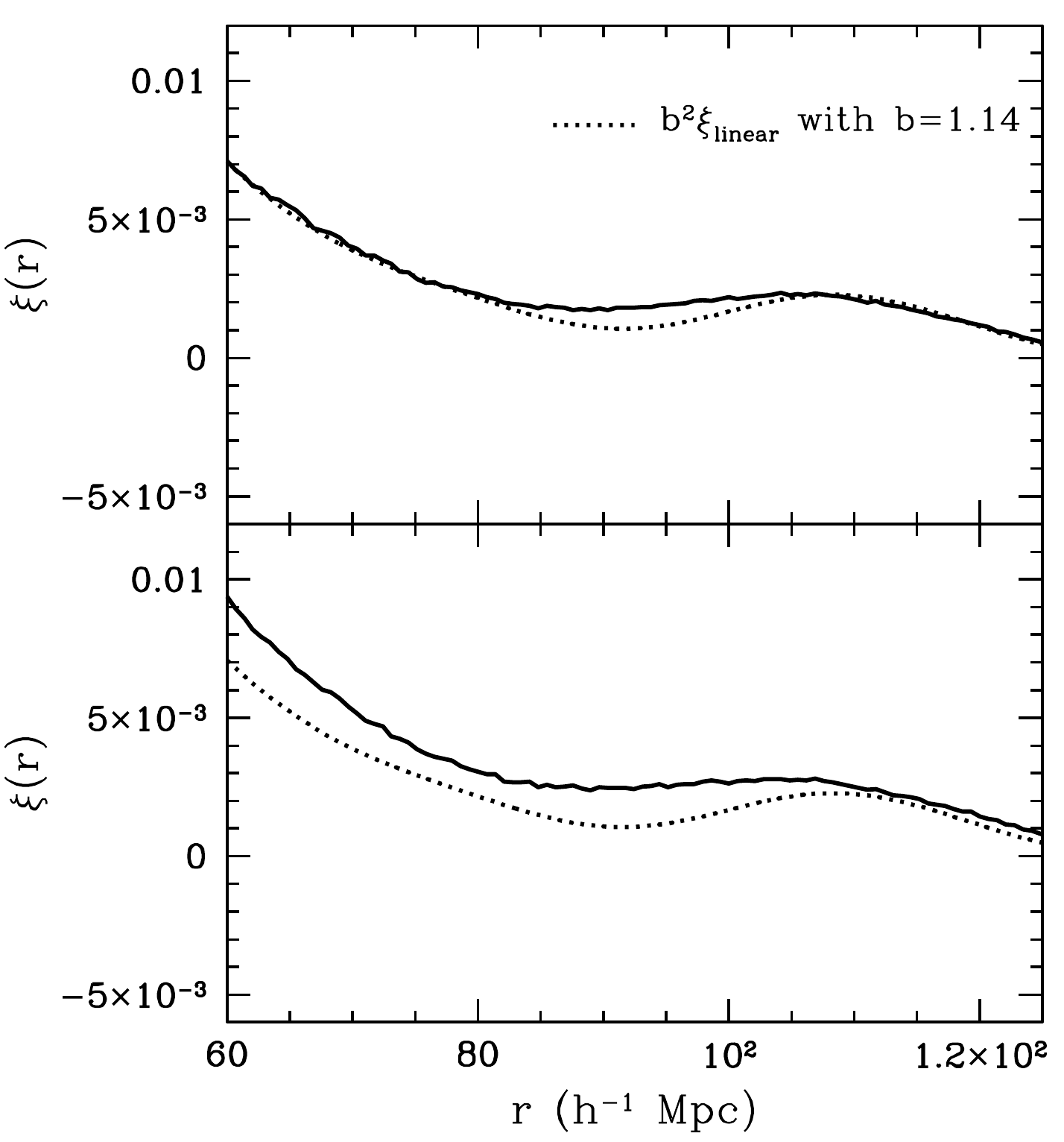} 
  \caption{Correlation function averaged over the directional cosine. 
    The  thick solid  line is  the averaged  value of  the correlation
    functions  in real  space  ({\it top})  and  redshift space  ({\it
      bottom panel}). 
    The dotted line is the linear prediction with bias $b=1.14$.
    \label{baoavg}
  }
\end{figure}

Figure~\ref{baoavg} shows  the average  correlation function  over the
directional cosine, 
\begin{equation}
\xi(r) \equiv \int_0^1 \xi(r,\mu) {\rm d}\mu .
\end{equation}
In both real and redshift spaces, the BAO peak from the HR4 is broadened
and shifted toward small scales compared to a simple estimation of the
linear correlation function of biased objects 
\begin{equation}
\xi_{\rm linear, bias}(r) = b^2 \xi_{\rm linear}(r),
\end{equation}
where $b = 1.14$ is the bias factor.
This  is due  to  {the} nonlinear  gravitational evolution  of
galaxies.  
In    redshift   space{,}    the   BAO    peak   is    further
broadened{,} and  it is hard  to clearly find the  position of
the BAO peak. 


\section{Mass Accretion History}

\begin{figure}[tp]
\centering
\includegraphics[width=8.5cm]{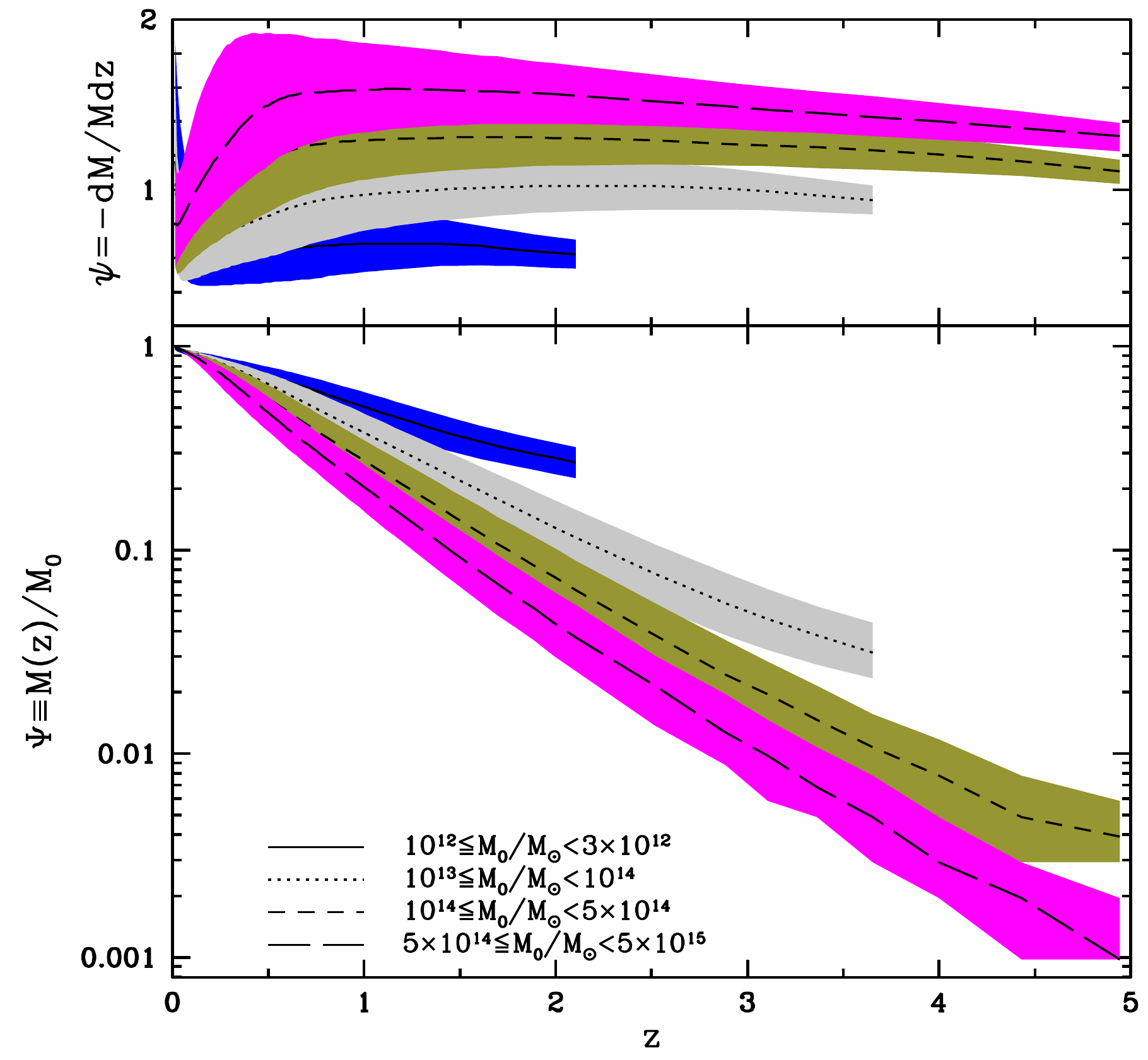} 
\caption{Evolution of halo mass with redshift for several mass samples.
In the top panel, we show the change of $\psi$ with redshift,
while $\Psi$ is shown in the bottom panel.
Lines and shaded regions mark the mean and $1\sigma$ distributions of mass history.
For each sample, we cut the data below the mass resolution of the simulation.
\label{mergingdist}
}
\end{figure}

We use  merger trees to study  the mass accretion history  of halos in
several mass samples. 
We define the mass accumulation history as
\begin{equation}
\Psi(M_0,z) \equiv {M(z) \over M_0},
\end{equation}
where $M_0$ is the final halo mass at $z=0$.  
The  half-mass  epoch   ($z_{1/2}$)  is  defined  as   the  time  when
$M(z_{1/2}) = M_0/2$. 
We measure the evolution of the halo mass along the major descendant trees
and show the results in Figure~\ref{mergingdist}. 
It can  be seen that the  half-mass redshift tends to  decrease as the
final halo mass increases.  
For example, {low-mass} halos with $10^{12} \le M_0/h^{-1}{\rm
  M_\odot} <  3 \times 10^{12}$ {tend  to} have {their
  half-mass around} $z_{1/2} \simeq 1$ on average, 
while {more  massive} halos with $10^{14}  \leq M_0/h^{-1}{\rm
  M_\odot} < 5 \times 10^{14}$ {tend to} have {a later
  half-mass epoch} $z_{1/2} \simeq 0.5$. 
This result is  consistent with the observations  that galaxy clusters
formed  relatively recently  (in terms  of  the epoch  when a  cluster
obtains half of the current mass) while individual satellite galaxies
seem to form at relatively higher redshift.

We empirically fit the log-linear function of redshift to $\Psi(z)$ as
\begin{equation}
\Psi(M_0,z) = \exp \left[-\psi(M_0) z \right],
\label{maccrete}
\end{equation}
and found a best-fit relation as
\begin{equation}
\psi(M_0) \simeq 0.32 \log_{10} \left({ M_0 \over 10^{12}~h^{-1}{\rm M_\odot} }\right) + 0.56.
\end{equation} 
This  fitting function  reproduces the  distribution for  $10^{12} \le
M/h^{-1}{\rm M_\odot} <5\times 10^{15}$ quite well at an early epoch ($z
\gtrsim 0.7$). 
On the other hand, at  low redshifts ($z\lesssim 0.7$) the accelerated
expansion  driven  by   dark  energy  begins   to  overpower  the
gravitational attraction and, therefore, halo mergers are suppressed.  
The sharp  increase of $\psi$  near the current  epoch is caused  by a
numerical noise.  
Note  that \citet{dekel13}  also  found an  exponential  form of  mass
growth, although  their mass accretion  rate ($\psi(10^{12}~h^{-1}{\rm
  M_\odot})    =    0.76$)    is    slightly    higher    than    ours
($\psi(10^{12}~h^{-1}{\rm M_\odot}) = 0.56$).

We want  to point out that  the specific mass accretion  rate per unit
redshift interval, defined as 
\begin{equation}
\left|{{\rm d}M\over M {\rm d}z}\right|  = \left|{M_0\over M} {{\rm d}
    \Psi \over {\rm d}z}\right| \approx \psi(M_0), 
\end{equation}
is roughly constant with redshift and depends only on the current sample mass.
Then the specific mass accretion rate per unit physical time can be calculated as
\begin{eqnarray}
\Upsilon_M (M_0,z) &\equiv&
\left|{{\rm d}M\over M {\rm d}t}\right| \\
&=&  { \psi(M_0) \over H_0} {E(z)\over 1+z }.
\end{eqnarray}

We now introduce a star formation  efficiency, which is defined as the
ratio of mass accretion rates between  the stellar and total masses of
halos as 
\begin{equation}
b_{\star} (M_0,z)  \equiv {\Upsilon_\star \over \Upsilon_M},
\end{equation}
where $\Upsilon_\star \equiv {\rm d}M_\star/M_\star {\rm d}t$ is the specific stellar mass accretion rate.
As a simple case, we assume that  the stellar mass evolution of a halo
is fully determined by the evolution of its total mass. 
In  this case,  the spectral  indices of  stellar mass-to-total  mass,
stellar  mass  accretion  rate-to-stellar  mass,  and  star  formation
efficiency-to-total mass, respectively defined as 
\begin{eqnarray}
\gamma(M_0, z) &\equiv& { {\rm d}\ln M_\star \over {\rm d}\ln M} \\
\beta(M_0, z) &\equiv& { {\rm d}\ln \Upsilon_\star \over {\rm d}\ln M_\star} \\
\epsilon(M_0, z) &\equiv& { {\rm d}\ln b_\star \over {\rm d}\ln M},
\end{eqnarray}
are fully determined by the redshift and the final halo mass.
By applying the galaxy-subhalo  correspondence model to relate between
halo mass and galaxy luminosity, \cite{kim08} showed that 
stellar luminosity (or stellar mass if a constant $M_\star/L_\star$ is
assumed)  shows a  good relation  to the  halo mass  with a  power-law
index $\gamma  \sim 0.5$  for the  SDSS main  galaxy sample  when $M
\gtrsim 5\times 10^{11} h^{-1}{\rm M_\odot}$.  
A similar slope was reported by \cite{kravtsov14} from the BCG samples.
On  the other  hand, \cite{abramson14}  reported $\beta\sim  -0.3$ for
SDSS DR7 galaxies with $9.5 \le \log_{10} (M_\star/h^{-1} M_\odot) \le
11.5$.

\begin{figure}[tp]
\centering
\includegraphics[width=8.4cm]{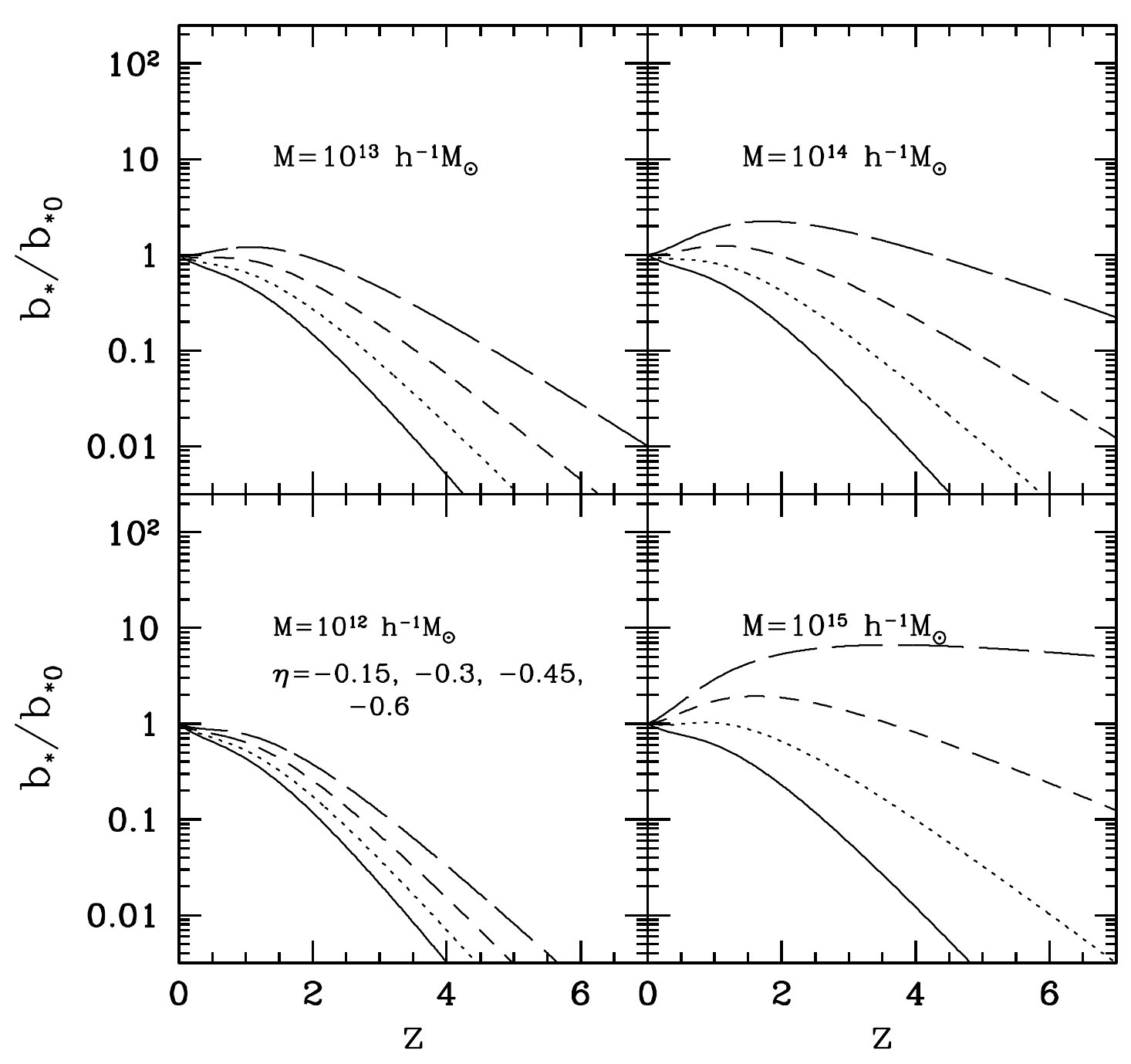} 
\caption{Relative star  formation efficiency  scaled with  the current
  efficiency, $b_{\star0}$. 
Clockwise   from   the  lower-left   panel,   the   halo  masses   are
$M_0=10^{12}~h^{-1}{\rm M_\odot}$,  $M_0=10^{12}~h^{-1}{\rm M_\odot}$,
$M_0=10^{13}~h^{-1}{\rm    M_\odot}$,   and    $M_0=10^{15}~h^{-1}{\rm
  M_\odot}$, respectively. 
In  the  legend,  we  list  the  values  of  $\eta$  from  the  bottom
curve.  
\label{sbias} 
}
\end{figure}

The spectral  index of the  stellar mass accretion  rate-to-total mass
can be expressed as a combination of the above spectral indices: 
\begin{eqnarray}
\eta(M_0, z) &\equiv& { {\rm d}\ln \Upsilon_\star \over {\rm d}\ln M} \\
&=& \beta(M_0, z) \gamma(M_0, z) \\
&=& \epsilon(M_0, z) + {1\over \psi (M_0)}\left[
 {{\rm d}\ln E(z) \over   {\rm d}z} -1 \right],
\end{eqnarray}
where $E(z)$ was defined in Equation~(\ref{eq:Ez}).
The effect of the parameters on the relative star formation efficiency is
shown in Figure~\ref{sbias}.  
As can be seen from the figure, the relative star formation efficiency
is higher{,} or the $\eta$ is getting smaller for more massive
halos.

\section{Summary\label{sec:con}}

{We ran a new cosmological $N$-body simulation called the Horizon Run 4 (HR4) simulation.
By adopting a standard $\Lambda$CDM cosmology in concordance with WMAP
5-year observations, the  HR4 simulates a periodic cubic box  of a side
length, $L_{\rm box} = 3150 h^{-1}{\rm Mpc}$ with $6300^3$ particles. 
With its wide range of mass and length scales, the HR4 can}
 provide the cosmology community with
a competitive data set for the study of cosmological models
and galaxy formation in the context of large-scale environments. 

{The main products of the HR4 are as follows.
First, we  saved the snapshot data  of the particles within  the whole
simulation box at 12 different redshifts from $z = 4$ to 0. 
We also built a past lightcone space data of particles that covers the
all-sky up to $z \simeq 1.5$. 
They}  can  be  used  to  study the  evolution  of  the  gravitational
potential and the  genus topology as well as  large-scale weak lensing
analysis. 
Moreover, we  constructed the  merger trees of  Friend-of-Friend halos
from  $z =  12$  to 0  with their  gravitationally  most bound  member
particles.  
They can be used to study  galaxy formation and bridge the gap between
theoretical models and observed galaxy distributions.

{We  tested   the  HR4  in  various   aspects,  including  the
  mass/shape/spin  distributions of  FoF halos,  two-point correlation
  functions of  physically self-bound subhalos, and  mass evolution of
  FoF halos.  
The results of our test are summarized as follows:} 
\begin{enumerate}
\item We found that  the abundance of massive FoF halos  in the HR4 is
  substantially  different from  various  fitting  functions given  in
  {the}  previous literature.   We also  found strong  evidence for  a
  redshift dependence of the mass 
function.  
We proposed  a new fitting  formula of the multiplicity  function that
reproduces the redshift changes of amplitude and shape of multiplicity
functions within about 5 \% errors. 

\item We confirmed the finding of previous studies that FoF halos tend
  to rotate around the minor axis. 

\item The  two-point correlation  function measured  in real  space is
  isotropic. 
However, due to the non-linear  evolution of galaxies, the location of
the  baryonic acoustic  oscillation  peak is  shifted toward  smaller
  scale than the prediction from the linear correlation function.  
On the other hand, in redshift space  the BAO peak can be seen only in
the two-point correlation function  along the perpendicular direction,
with a much{-}broadened width and increased height.  
We emphasize{d} that it is important to use massive simulation data to
study  the non-linear  evolution of  BAO features  and the  connection
between observations and cosmological models.  

\item  We found  that more  massive halos  tend to  have steeper  mass
  histories, and the  mass accretion rate per unit  redshift is roughly
  constant during early epoch before dark energy domination. 
By  adopting simple  power-law models  for the  stellar mass  and star
formation  efficiency, we  found that  massive  halos tend  to have  a
higher star formation efficiency. 

\end{enumerate}

{All aforementioned main products of  the HR4 are available at
  \url{http://sdss.kias.re.kr/astro/Horizon-Run4/}.} 


\acknowledgments

This         work         was         supported         by         the
Supercomputing         Center/Korea
Institute  of   Science  and  Technology   {Information  with}
supercomputing      resources     including      technical     support
(KSC-2013-G2-003).  
The authors  thank Korea  Institute for  Advanced Study  for providing
computing resources  (KIAS Center  for Advanced Computation)  for this
work.  
The authors also  thank the referee, Graziano Rossi,  for the thorough
review and constructive suggestions that lead to an improvement of the
paper. 


\end{document}